\documentclass[letterpaper,11pt]{article}
\usepackage{natbib,amsmath,amsfonts}

\usepackage[hyperindex,breaklinks]{hyperref}

\DeclareMathOperator*{\plim}{plim}
\DeclareMathOperator*{\argmax}{argmax}

\newcommand{\Ep}{{\mathrm{E}}}
\newcommand{\En}{{\mathbb{E}_n}}
\newcommand{\Gn}{{\mathbb{G}_n}}
\newcommand{\ENT}{{\mathbb{E}_{NT}}}
\newcommand{\EN}{{\mathbb{E}_N}}
\newcommand{\ET}{{\mathbb{E}_T}}
\newcommand{\GNT}{{\mathbb{G}_{NT}}}
\newcommand{\GN}{\mathbb{G}_N}
\newcommand{\GT}{\mathbb{G}_T}

\begin{document}

\title{\bf Fixed Effect Estimation of Large T \\ Panel Data Models\thanks{
         \scriptsize
We thank
St\'{e}phane Bonhomme,
Geert Dhaene,
and
Koen Jochmans for helpful comments, and
 we are grateful to Siyi Luo for proofreading, and to Mario Cruz-Gonzalez for helping with the Monte Carlo simulations.
Financial support from the National Science Foundation, 
and the Economic and Social Research Council through the ESRC Centre for Microdata Methods and Practice grant RES-589-28-0001,
and the European Research Council grant ERC-2014-CoG-646917-ROMIA
 is gratefully   acknowledged. 
}}

\author{\setcounter{footnote}{2} Iv\'{a}n Fern\'{a}ndez-Val\footnote{
     Department of Economics, Boston University,
     270 Bay State Road,
     Boston, MA 02215-1403, USA.
     Email:~{\tt ivanf@bu.edu}
      }
   \and Martin Weidner\footnote{
                   Department of Economics,
                   University College London,
                   Gower Street,
                   London WC1E~6BT,
                   UK,
                    and CeMMAP.
                   Email:~{\tt m.weidner@ucl.ac.uk}
                   }
                   }

\date{\today}
\maketitle

\bigskip
\bigskip

\begin{abstract}
This article reviews recent advances in fixed effect estimation of panel data models for
long panels, where the number of  time periods is relatively large.
We focus on semiparametric models with unobserved individual and time effects, where the distribution  
of the outcome variable conditional on covariates and unobserved effects is specified parametrically, while the distribution of the unobserved effects is
left unrestricted. Compared to existing reviews on long panels (\citealp{ArellanoHahn2007};
a section in \citealp{ArellanoBonhomme2011}) we  discuss
models with both individual and time effects, split-panel Jackknife bias corrections, 
unbalanced panels, distribution and quantile effects, and
other extensions. Understanding and correcting the incidental parameter bias caused by the
estimation of many fixed effects is our main focus,
and the unifying theme  is that the order of this bias
is given by the simple formula $p/n$ for all models discussed, with $p$  the number of estimated parameters
and $n$ the total sample size.
\end{abstract}

{\footnotesize

}

\newpage

\section{INTRODUCTION}
 
One of the main advantages of panel data over cross-sectional or time series data is the possibility of accounting for multiple sources of unobserved heterogeneity.
This is accomplished by including individual and time unobserved effects into the model, which control for
any unobserved covariate that is either time or cross-sectional invariant. In this article we consider fixed effect approaches to panel data models, where no distributional assumption on the unobserved effects is imposed,  therefore allowing the unobserved effects to be arbitrarily related with the observed covariates. Fixed effects contrast with random effects approaches that impose restrictions on the distribution of the unobserved effects conditional on the observed covariates. 
We refer to  \citet*{Arellano2003}, \citet*{Baltagi2008}, \citet*{Hsiao2015}, and \citet*{Wooldridge2010} for modern textbook treatments on the difference between fixed and random effects.

We consider semiparametric models that specify parametrically some characteristic of the distribution of the 
outcome variable of interest conditional on the observed covariates and unobserved effects. Examples include linear and nonlinear regression models for the conditional expectation, and  generalized linear, distribution and quantile regression models  for the conditional distribution.  Generalized linear models include the most commonly used nonlinear models such as probit, logit, Poisson, negative binomial, proportional hazard, and tobit models. The nonparametric or unspecified part of the model consists of the  joint distribution of the unobserved effects, exogenous covariates and initial conditions.  We do not review  nonparametric panel models such as the nonseparable models considered in  \citet*{AltonjiMatzkin2005}, \citet*{Evdokimov2010}, \citet*{GrahamPowell2012}, \citet*{HoderleinWhite2012}, \citet*{CFHN13}, \citet*{CFHHN15}, \citet*{CFN17}, \citet*{Freyberger2017}, 
 and \citet*{Torgovitsky2016},
   among others. We refer to \citet*{Matzkin2007}
  for an excellent survey on nonparametric identification of nonseparable models including results for panel data.

We analyze the properties of fixed effects estimators of model parameters. These estimators treat the unobserved effects as parameters to be estimated. In linear models with strictly exogenous covariates and individual effects, fixed effects is numerically equivalent to the within-group estimator that removes the individual effects by taking differences within each individual. We also analyze fixed effects estimation of average partial effects (APEs), which are averages of functions of the data, parameters and unobserved effects. APEs are often the quantities of interest in nonlinear models because they correspond to marginal effects of the covariates in some characteristic of the distribution of the outcome conditional on covariates and unobserved effects, averaged over the observed and unobserved heterogeneity. The main challenge with fixed effects estimators is to deal with the incidental parameter problem coming from the estimation of the unobserved effects. The estimated unobserved effects, usually called fixed effects, can be very noisy because there are few observations that are informative about them.  Section~\ref{sec:bias} describes the incidental parameter problem in detail.

Motivated by the first panel data sets available, the econometric theory was initially designed for panels where the number of time periods, $T$, is small compared to the number of cross-sectional units, $N$. This theory analyzed identification, estimation and inference using a fixed-$T$ asymptotic approximation that considers sequences of panels with increasing $N$ and fixed $T$.  Consistent approaches under this approximation usually consist of  removing the unobserved effects from the model by some smart transformation such as differencing or conditioning on sufficient statistics. Examples include \citet*{Mundlak1978}, \citet*{ArellanoBond1991}
and \citet*{Moreira:2009}
 for linear models;  \citet*{Cox1958}, \citet*{Rasch1960},
\citet*{Andersen1970},  \citet*{Chamberlain1980}, \citet*{Manski1987}, \citet*{Horowitz1992},
 \citet*{HonoreKyriazidou2000}, and \citet*{ArellanoCarrasco2003}
for static and dynamic binary choice models; \citet*{HausmanHallGriliches1984} for static Poisson models; 
\citet*{Honore1992} for tobit models;
\citet*{Kyriazidou1997,Kyriazidou2001} for sample selection models; and
\citet*{Muris2016} and \citet*{BotosaruMuris2017} for ordered logit and probit models;
 \citet*{Bonhomme2012} provided a unifying scheme for many of those methods.

The fixed-$T$ approach has encountered several limitations. First,  transformations to remove the unobserved effects are not available for some models such as probit and quantile regression. This issue gets  worse in models with unobserved effects in multiple dimensions, where even if  sufficient statistics exist conditioning on them can be computationally challenging \citep*{hirjietal87, Charbonneau2014}. Second, differencing over time in dynamic linear models does not work well when the data are very persistent because it removes most of the information about the parameters. Third, removing the unobserved effects from the model precludes the estimation of APEs in nonlinear models. Fourth, under fixed-$T$ asymptotics the incidental parameter problem of fixed effects estimation is a consistency problem, which is difficult to tackle. Fifth, some models are not $\sqrt{N}$-estimable or not even point identified  under the fixed-$T$ approximation, see , e.g.,  \citet*{Chamberlain2010}, \citet*{HonoreTamer2006}, \citet*{CFHN13}, \citet*{ShiShumSong2014}, and \citet*{PakesPorter2013}. 

All the limitations of the fixed-$T$ approximation mentioned in the previous paragraph, together with the recent availability of long panel datasets, has led to an alternative asymptotic approximation to panel data that considers sequences of panels where both $N$ and $T$ increase. This large-$T$ approximation, developed by \citet*{Phillips:1999p733} for linear models,  deals with some of the shortcomings of the fixed-$T$ approximation. For example, most models are point identified with large-$T$ and the incidental parameter problem of fixed effects estimation becomes an asymptotic bias problem that is easier to tackle.  Bias corrections have been developed to provide improved fixed effects  estimators of model parameters and APEs.  This large-$T$ approximation can also be applied naturally to other types of data with a grouping structure similar to panel data such as network and trade data.

The two alternative asymptotic approximations to panel data should
be viewed as complements rather than substitutes. For example,  
if a fixed $T$ consistent estimator is available for the particular parameter under consideration,
then it generally should be used. Otherwise
it is likely that the model is not point identified for fixed-$T$. Then, one can either consider bounding the parameter if  $T$ is small, or drawing on a large-$T$ consistent estimator if $T$ is moderately large with respect to $N$.
\citet*{HonoreTamer2006} and \citet*{CFHN13} showed that in some binary response models the bounds for parameters and APEs can be very tight and shrink very rapidly with $T$. Thus, unless $T$ is very small, 
one can safely ignore  that the model might only be set-identified
and  use large $T$ consistent estimators. Alternatively, one might entertain (correlated) random effects approaches that achieve fixed-$T$ point identification by imposing restrictions on the distribution of the unobserved effects.

This review focuses solely on fixed effects estimators of semiparametric panel models analyzed under a large-$T$ asymptotic approximation. Available survey articles on panel data estimation under fixed-$T$ approximations include   \citet*{Chamberlain1984}, \citet*{ArellanoHonore2001}, \citet*{Honore2002}, \citet*{ArellanoBonhomme2011}, and \citet*{ArellanoBonhomme2017}.  We concentrate mainly on the developments  since \citet*{ArellanoHahn2007}, which previously reviewed the large-$T$ approach to panel data. This field has grown sufficiently fast in recent years to justify another review. In particular, \citet*{ArellanoHahn2007} considered almost exclusively models with only unobserved individual effects, whereas we consider  models with unobserved effects in multiple dimensions such as individual and time effects. 
We also review
the estimation of distributional and quantile effects with panel data, which were not discussed in \citet*{ArellanoHahn2007}. Our review relies heavily upon the important foundational work in this literature by 
 \citet*{NeymanScott1948}, \citet*{Nickell81}, \citet*{coxreid87}, \citet*{kiviet95},  \citet*{Phillips:1999p733}, \citet*{Lancaster2000},  \citet*{Hahn:2002p717},  \citet*{Lancaster:2002p875},  \citet*{Woutersen:2002p3683}, \citet*{arellano03}, \citet*{AlvarezArellano2003},  \citet*{li2003efficiency},  \citet*{Hahn:2004p882},  \citet*{Carro:2007p3601},
 \citet*{FernandezVal:2009p3313},  \citet*{HahnKuersteiner2011},  \citet*{gmm13}
 and \citet*{DhaeneJochmans2015}, among others.

\paragraph*{Outline:} In Section~\ref{sec:bias} we describe the bias
of maximum likelihood estimators (MLEs) in cross-sectional samples, which allows us to establish some 
key results  in a simplified setting. Those results
are then generalized to panel data models in Section~\ref{sec:SPM}, where we discuss the
incidental parameter bias and bias correction methods for fixed effected estimators of model parameters 
and average partial effects. A short heuristic derivation of the main formulas in Section~\ref{sec:SPM}
is provided in the appendix. Section~\ref{sec:extensions} discusses
unbalanced panels, multivariate fixed effects, distributional and quantile effects, and some further extensions
of the methods, before we conclude in Section~\ref{sec:conclusions}.

\paragraph*{Notation:} 
In the following we denote by $\Ep$ the expectation with respect to the parametric part of the model.
Averages over cross-sectional and time-series samples are denoted by
$\En = n^{-1} \sum_{j=1}^n$, 
$ \mathbb{E}_N  =  N^{-1} \sum_{i=1}^N$,
$ \mathbb{E}_T  = T^{-1} \sum_{t=1}^T$,
 and $ \mathbb{E}_{NT} = (NT)^{-1} \sum_{i=1}^N \sum_{t=1}^T$.
 For example, we write $\mathbb{E}_n  \ell_{j}(\theta) $ for $n^{-1} \sum_{j=1}^n \ell_{j}(\theta) $ in the next section.
We also define  $\Gn =  \sqrt{n} \left( \En  - \Ep \right) $,
$\GN  =  \sqrt{N} \left( \EN  -  \EN \Ep \right)$,
$\GT  =  \sqrt{T} \left( \ET  -  \ET \Ep \right)$,
and
$\GNT  =  \sqrt{NT} \left( \ENT  -  \ENT \Ep \right)$.
The notations $\En$ and $\Gn$ are used in Section~\ref{sec:bias} for random samples, whereas $ \mathbb{E}_N$, $ \mathbb{E}_T$,  $\GN$, $\GT$ and $\GNT$ are used in  Section~\ref{sec:SPM} for nonrandom panel samples, which explains the additional sample averages before the expectations $\Ep$. 
For example  $\GNT  \ell_{it} = \sqrt{NT}  \ENT \left(  \ell_{it} - \Ep \ell_{it} \right)$.
In Section~\ref{sec:bias} we often just write a bar to denote the expectation $\Ep$, e.g.
 $\overline \ell_{it} = \Ep \ell_{it} $.
Vector-valued  variables are always arranged as column-vectors in the following. Thus,
we  write   $\theta=(\beta, \alpha, \gamma)$ to form the column-vector $\theta$ as the concatenation of the column-vectors $\beta$, $\alpha$ and $\gamma$. Vector and matrix transposition is denoted by a prime. We use $\overset{a}{\sim}$ to denote asymptotic approximation to the distribution. For example, if $\sqrt{n}(\widehat \beta - \beta_0 - B/a_n) \to_d \mathcal{N}(0,V)$ for some sequence $a_n$ such that $a_n \to \infty$ as $n \to \infty$, then we write 
$\widehat \beta - \beta_0 \overset{a}{\sim} \mathcal{N}(B/a_n, V/n)$.

\section{THE INCIDENTAL PARAMETER PROBLEM}\label{sec:bias}
Incidental parameters are nuisance parameters whose dimension grows with the sample size. \citet{NeymanScott1948} showed that maximum likelihood estimators (MLEs) can be asymptotically biased in models with incidental parameters. Here we provide a  derivation based on a second-order asymptotic expansion that the order of the bias of the MLEs in random samples is
\begin{align}
     {\rm bias}  \sim  \frac{p} {n},
    \label{BiasOrder}
\end{align}
where $p$ is the number of parameters and $n$ is the sample size. This order corresponds with the inverse of the number of observations per parameter. In the following sections we will see that the order of the bias of fixed effects estimators of parameters and APEs can also be obtained using this simple heuristic formula in more complex panel data samples.

\subsection{Bias of Maximum Likelihood Estimators}
\label{sec:IIDexpansion}

Given a random sample $\{z_j : 1 \leq j \leq n\}$ of a random variable $z$ that has density $f(z,\theta_0)$ with respect to some dominating measure, let  
\begin{align}\label{eq:mest}
     \widehat \theta &=  \argmax_{\theta \in \mathbb{R}^p}  \,  \mathbb{E}_n  \ell_{j}(\theta) ,
     &
     \ell_{j}(\theta) &:=  \log f(z_j,\theta),
\end{align}
be the maximum likelihood estimator (MLE) of the value of the $p$-dimensional parameter $\theta$ that is identified by the population program
\begin{equation}\label{eq:mest2}
\theta_0 = \argmax_{\theta \in \mathbb{R}^p} \Ep  \ell_{j}(\theta).
\end{equation}
In the following expansions we assume that $\theta_0$ is uniquely determined by the first order conditions of the program \eqref{eq:mest2}, $\theta \mapsto \ell_j(\theta)$ is a.s.\  differentiable to sufficient order, and other regularity conditions that allow us to bound remainder terms.  We also assume that $\widehat \theta$ is consistent, i.e., $\widehat \theta \to_P \theta_0$ under some norm.  Establishing this consistency might require delicate arguments in high dimensional settings where we let $p \to \infty$ as $n \to \infty$. We denote derivatives of $\theta \mapsto \ell_j(\theta)$ using superscripts $\theta$, that is, 
$\ell^{\theta}_{j}(\theta)$ is the $p$-dimensional gradient, $\ell^{\theta \theta}_{j}(\theta)$ is the $p \times p$ Hessian matrix, and, for  $k \in \{1,\ldots,p\}$, $\ell^{\theta \theta \theta_k}_{j}(\theta)$ is the $p \times p$
matrix obtained by taking the partial derivative of the Hessian with respect to $\theta_k$.
We drop the argument $\theta$ from all the functions whenever they are evaluated at $\theta_0$, e.g.
$\ell^{\theta}_{j} := \ell^{\theta}_{j}(\theta_0)$.

By consistency, the asymptotic properties
of $\widehat \theta$ are governed by the local properties of $\ell_{j}(\theta)$ around $\theta_0$.
A second order Taylor expansion of the first order condition of \eqref{eq:mest} around $\theta=\theta_0$ yields
\begin{align*}
    0 = \En  \ell^\theta_{j}(\widehat \theta) \approx \mathbb{E}_n  \ell^\theta_{j}  
      +  \mathbb{E}_n  \ell^{\theta \theta}_{j}   \, (\widehat \theta - \theta_0)
      + \frac 1 2  \sum_{k=1}^p  \mathbb{E}_n  \ell^{\theta \theta \theta_k}_{j}   \, (\widehat \theta - \theta_0) (\widehat \theta_k - \theta_{0,k}).
\end{align*}
We have $ \mathbb{E}_n  \ell^\theta_{j}   =  n^{-1/2}  \mathbb{G}_n  \ell^\theta_{j} $, because $ \Ep \ell^\theta_{j} = 0$ by the first order conditions of the population program \eqref{eq:mest2}. Decomposing $  \mathbb{E}_n  \ell^{\theta \theta}_{j}  =   \Ep  \ell^{\theta \theta}_{j}  +   n^{-1/2}  \mathbb{G}_n  \ell^{\theta \theta}_{j} $ and  $  \mathbb{E}_n  \ell^{\theta \theta \theta_k}_{j}  =   \Ep  \ell^{\theta \theta \theta_k}_{j}  +   n^{-1/2}  \mathbb{G}_n  \ell^{\theta \theta \theta_k}_{j} $, and ignoring the term that includes $\mathbb{G}_n  \ell^{\theta \theta \theta_k}_{j}$  because it is of small enough order, gives
\begin{align}\label{eq:exp}
      n^{-1/2}  \mathbb{G}_n  \ell^\theta_{j}   
      +  \Ep  \ell^{\theta \theta}_{j}   \, (\widehat \theta - \theta_0)
      +  n^{-1/2}  \mathbb{G}_n  \ell^{\theta \theta}_{j}   \, (\widehat \theta - \theta_0)
      +  \frac 1 2 \sum_{k=1}^p  \Ep  \ell^{\theta \theta \theta_k}_{j}   \, (\widehat \theta - \theta_0) (\widehat \theta_k - \theta_{0,k})
    \approx 0 .
\end{align}
The leading two terms of \eqref{eq:exp} give the standard {\it first-order} approximation  
$\widehat \theta - \theta_0 \approx  n^{-1/2} \psi_1$,
where $\psi_1 := - \left( \Ep  \ell^{\theta \theta}_{j}  \right)^{-1}  \mathbb{G}_n  \ell^\theta_{j}   $ is the influence function.
Plugging this first-order approximation into the third and fourth term of \eqref{eq:exp} and dropping low order terms yields the
{\it second-order} approximation 
\begin{align}
  \widehat \theta - \theta_0 &\approx  n^{-1/2}  \psi_1 + n^{-1}  \psi_2  ,
  \label{SecondOrderExpansionIID}
\end{align}  
where
\begin{align*}
    \psi_2 &:= -   \left( \Ep  \ell^{\theta \theta}_{j}  \right)^{-1} \mathbb{G}_n  \ell^{\theta \theta}_{j}   \, \psi_1
                     - \frac 1 2  \sum_{k=1}^p
                      \left( \Ep  \ell^{\theta \theta}_{j}  \right)^{-1}    \Ep  \ell^{\theta \theta \theta_k}_{j}   \, \psi_1 \psi_{1,k} .
\end{align*}
We now analyze the properties of the components of $\psi_1 = (\psi_{1,1},\ldots, \psi_{1,p})$ and $\psi_2 = (\psi_{2,1},\ldots, \psi_{2,p})$.   For each $k \in \{1,\ldots,p\}$, $\psi_{1,k}$ has zero mean,  finite variance and asymptotic normal distribution under standard assumptions. Thus, $\psi_1$ is a variance term that does not contribute to the bias of $\widehat \theta$.  To determine the mean of the quadratic term  $\psi_{2,k}$, we assume for simplicity that  $\Ep  \ell^{\theta \theta}_{j} $ is a diagonal matrix.\footnote{This simplification is without loss of generality because we can diagonalize the Hessian  by reparametrizing the model.} Then,
\begin{align}
    \Ep  \psi_{2,k}
      &= 
      \sum_{l=1}^p   
   \frac{\Ep\left(  \ell^{\theta_k \theta_l}_{j} \ell^{\theta_l}_{j}  
                     +  \frac 1 2  \,  \ell^{\theta_k \theta_l \theta_l}_{j}   \right) } 
                               { \Ep  \ell^{\theta_k \theta_k}_{j}  \,\Ep  \ell^{\theta_l \theta_l}_{j} }
  ,
      \label{BiasIIDcase}
\end{align}
where  we use the information equality 
$  \Ep  \ell^{\theta_l}_{j}  \ell^{\theta_m}_{j} =  -  \Ep  \ell^{\theta_l \theta_m}_{j}$ to simplify the expression.
This result shows that $\Ep \psi_{2,k}$ is proportional to $p$, because of the sum over $l = 1,\ldots,p$,
and therefore
$$
\Ep  \widehat \theta - \theta_0  \approx  \underbrace{ n^{-1/2}   \Ep \psi_1 }_{=0} + n^{-1}  \Ep \psi_2 \sim  n^{-1} p,
$$
which verifies the order of the bias in  \eqref{BiasOrder}. The term $\psi_2$ also has  variance. However, the  order of the variance of the term $n^{-1} \psi_2$  is lower than the order of the variance of the term $n^{-1/2} \psi_1$ in the expansion \eqref{SecondOrderExpansionIID}.

\subsection{Consequences for Inference}\label{subsec:inf}
The asymptotic bias of the MLE has consequences for  inference. Confidence intervals based on the first-order asymptotic distribution might severely  undercover even in large samples if the sample size is not sufficiently large relative to the dimension of the parameters, more precisely when $n = O(p^2)$. To see this, let 
$\rho'\theta$, $\rho \in \mathbb{R}^p$, be the inferential object of interest,
and also let $\Ep \psi_2/p = B = (B_1, \ldots, B_p)$. By standard theory for MLE,    $\rho'\psi_1 \to_d \mathcal{N}(0,\rho'\Omega\rho)$ with $\Omega = - \left( \Ep  \ell^{\theta \theta}_{j}  \right)^{-1}$, and $\rho'\psi_2/p \to_p \rho'B$, as $n\to \infty$. Then, as $n \to \infty$,
$$
\rho' \left(\widehat \theta - \theta_0 \right) \overset{a}{\sim} \mathcal{N}\left(   \frac{p}{ n} \, \rho'B,  \;  \frac 1 n \, \rho'\Omega\rho \right).
$$
Consider $\rho=e_k$, the unit vector with a one in position $k$. The standard asymptotic two-sided  $(1-\alpha)$-confidence interval for $\theta_k = e_k'\theta$ is 
$$
CI_{1-\alpha}(\theta_k) = \widehat \theta_k \pm z_{\alpha/2} \widehat \Omega^{1/2}_{kk}/\sqrt{n},
$$
where $\widehat \Omega_{kk}$ is a consistent estimator of $\Omega_{kk}$, the $(k,k)$ element of  $\Omega$, and $z_{\alpha/2}$ is the $(1-\alpha/2)$-quantile of the standard normal distribution.

Consider $p/\sqrt{n} \to \kappa$ as $n \to \infty$. Then, the coverage of $CI_{1-\alpha}(\theta_k)$ in large samples is
\begin{multline*}
\Pr(\theta_k \in CI_{1-\alpha}(\theta_k)) = \Pr(\sqrt{n} |\widehat \theta_k - \theta_k| \leq z_{\alpha/2} \widehat \Omega_{kk}^{1/2}) \to \Pr(|\mathcal{N}(\kappa B_k, \Omega_{kk})| \leq z_{\alpha/2}  \Omega_{kk}^{1/2}) \\ = \Phi(z_{\alpha/2}  - \kappa B_k/\Omega_{kk}^{1/2}) - \Phi(- z_{\alpha/2}  - \kappa B_k/\Omega_{kk}^{1/2}) < 1-\alpha
\end{multline*}
if $\kappa B_k \neq 0$. This coverage can be much lower than the nominal level $1-\alpha$. For example, if  $\kappa B_k/\Omega_{kk}^{1/2} = 1$, then the coverage probability of a 95\% interval is less than 83\% in large samples.

\section{SEMIPARAMETRIC PANEL MODELS}
\label{sec:SPM}

\subsection{Model} We observe the panel data set $\{(y_{it}, x_{it}): 1 \leq i \leq N, 1 \leq t \leq T \},$ for a scalar outcome variable of interest  $y_{it}$ and a vector of covariates $x_{it}$. The subscripts $i$ and $t$ index individuals and  time periods in traditional panels, but they might index other dimensions in more general data structures such as firms and industries or siblings and families. The observations are independent across $i$ and weakly dependent across $t$.\footnote{
Long panel models with cross-sectional dependence are discussed in \citet*{Pakel2014}.
}
 We consider the semiparametric model for each $i = 1,\ldots,N$: 
\begin{equation}\label{eq:spm}
y_{it}  \mid x^t_i, \alpha, \gamma   \sim f(\cdot \mid x_{it},\alpha_i, \gamma_t; \beta), \ \text{independently over} \ t = 1, \ldots, T,
\end{equation}
where  $x^t_i = (x_{i1}, \ldots, x_{it}),$  $\alpha = (\alpha_1, \ldots, \alpha_N)$, $\gamma = (\gamma_1, \ldots, \gamma_T)$,
$f$ is a known density with respect to some dominating measure, and $\beta$ is a $d_{\beta}$-vector of parameters.
The variables $\alpha_i$ and $\gamma_t$ are scalar unobserved individual and time effects that in economic applications capture individual heterogeneity and aggregate shocks, respectively.\footnote{We refer to \citet*{gmm13} for GMM panel models defined by moment conditions. We discuss models with multivariate individual and time effects in Section \ref{sec:MultivariateFE}.} The model is semiparametric because  it does not specify the distribution of these effects nor their relationship with the covariates. The conditional density $f$ represents the parametric part of the model.  The covariates $x_{it}$ are predetermined with respect to $y_{it}$ and might include lags of $y_{it}$ to accommodate dynamic models. If the covariates $x_{it}$ are strictly exogenous with respect to $y_{it}$, then $x^t_i$  can be replaced by $x^T_i=(x_{i1}, \dots, x_{iT})$ in the conditioning set. Strict exogeneity  rules out dynamics and more generally any feedback from the outcome to future values of the covariates, which might be restrictive in applications.

\paragraph*{Example (i)}
The simplest example of this model is the normal  linear model with additive individual and time effects $y_{it} = x_{it}'\delta + \alpha_i + \gamma_t + \varepsilon_{it},$ $\varepsilon_{it} \mid x^t_i, \alpha, \gamma   \sim \mathcal{N}(0,\sigma^2),$ where
\begin{equation}\label{eq:linear}
f(y \mid x_{it},\alpha_i, \gamma_t; \beta) = \frac{1}{\sqrt{2\pi \sigma^2}} \exp\left[-\frac{(y-x_{it}'\delta - \alpha_i - \gamma_t )^2}{2\sigma^2}\right], \quad \beta = (\delta,\sigma^2).
\end{equation}

\paragraph*{Example (ii)}
An example of a nonlinear model is the panel binary response single index model with additive unobserved individual and time effects:
$$y_{it} = 1(x_{it}'\beta + \alpha_i + \gamma_t \geq \varepsilon_{it}), \quad \varepsilon_{it} \mid x^t_i, \alpha, \gamma   \sim F_{\varepsilon},$$
where $F_{\varepsilon}$ is a known CDF such as the standard normal or logistic. In this case,
\begin{equation}\label{eq:probit}
f(y \mid x_{it},\alpha_i, \gamma_t; \beta) = F_{\varepsilon}(x_{it}'\beta + \alpha_i + \gamma_t )^y
\times [1-F_{\varepsilon}(x_{it}'\beta + \alpha_i + \gamma_t )]^{1-y}  \times 1(y \in \{0,1\}).
\end{equation} 

\paragraph*{Example (iii)}
Another example is a panel Poisson count response single index model with additive unobserved individual and time effects where, for $\lambda_{it} = \exp(x_{it}'\beta + \alpha_i + \gamma_t )$,
\begin{equation}\label{eq:poisson}
f(y \mid x_{it},\alpha_i, \gamma_t; \beta) = \frac{\lambda_{it}^{y} \exp(-\lambda_{it})}{y!} \; 1\left(y \in \{0,1,\ldots\}\right).
\end{equation}

\medskip

Let  $\beta_0$,  $\alpha_0 = (\alpha_{01}, \ldots, \alpha_{0N})$, and $\gamma_0 = (\gamma_{01}, \ldots, \gamma_{0T})$ denote the values of $\beta$, $\alpha$ and $\gamma$ that generate the data. We assume that these values are identified by the population conditional maximum likelihood program
\begin{align}\label{eq:fe-pops}
(\beta_0,\alpha_0,\gamma_0) & \in \argmax_{(\beta, \alpha, \gamma) \in \mathbb{R}^{d_{\beta}+N+T}} \ENT \bar \ell_{it}(\beta, \alpha_i, \gamma_t), \ &
      \ell_{it}(\beta, \alpha_i, \gamma_t) &=   \log f(y_{it} \mid x_{it}, \alpha_i, \gamma_t;  \beta) ,
\end{align}
where $\bar \ell_{it}(\beta, \alpha_i, \gamma_t) := \Ep  \ell_{it}(\beta, \alpha_i, \gamma_t)$, the expected conditional 
log-likelihood with respect to the parametric part of the model. 
This program can have multiple solutions for $\alpha_0$ and $\gamma_0$. For example, in models that are additively separable in $\alpha_i$ and $\gamma_t$ such as the single index models above, $\ell_{it}(\beta, \alpha_i, \gamma_t) = \ell_{it}(\beta, \alpha_i + \gamma_t)$, the location translation $\alpha_i + c$ and $\gamma_t -c$ for a constant $c$ does not change the objective function of the program. In this case  we assume that there exists a normalization that selects   $\alpha_0$ and $\gamma_0$ 
  (e.g. $\alpha_{01}=0$ or $\gamma_{01}=0$).

\subsection{Fixed Effects Estimator and Incidental Parameter Problem}

The fixed effects (FE) estimator  treats the realizations of $\alpha$ and $\gamma$ as parameters to be estimated. It is the solution to the sample conditional maximum likelihood program, for $\widehat \alpha = (\widehat \alpha_{1}, \ldots, \widehat \alpha_{N})$ and $\widehat \gamma = (\widehat \gamma_{1}, \ldots, \widehat \gamma_{T})$,
\begin{align}\label{eq:fe}
(\widehat \beta, \widehat \alpha,\widehat \gamma) & \in \argmax_{(\beta, \alpha, \gamma) \in \mathbb{R}^{d_{\beta}+N+T}} \ENT  \ell_{it}(\beta, \alpha_i, \gamma_t), 
\end{align}
which is the sample analog of \eqref{eq:fe-pops}. This program can also have multiple solutions for $\widehat \alpha$ and $\widehat \gamma$. In that case we adopt the same normalization as in the population program to select $\widehat \alpha$ and $\widehat \gamma$.  In the binary and count response examples, we can obtain the FE estimator using standard software routines including individual and time indicators for the unobserved effects. We discuss some computational aspects of the program \eqref{eq:fe}  in Section~\ref{subsec:com}.

The FE estimator suffers from the
incidental parameter problem  (\citealp{NeymanScott1948},
see also \citealp{Lancaster2000} for a review). The incidental parameters are the individual and time fixed effects which have dimensions $N$ and $T$, respectively.  Unless $T$ is large,
each individual fixed effect is very noisy because  only $T$ observations are informative about it. Symmetrically, unless $N$ is large, each time fixed effect is very noisy.   The noise in the fixed effects  generally contaminates the estimators of the other parameters. Exceptions include linear and Poisson models with strictly exogenous covariates where  it is possible to separate the estimation of the fixed effects from the other parameters.  
The asymptotic consequences of the incidental parameter problem depend on the approximation adopted. It is a consistency problem under a fixed-$T$ or fixed-$N$ approximation, whereas it becomes an asymptotic bias problem under a large-$T$ and large-$N$ approximation, which we adopt here. 

\subsection{Asymptotic Bias}
\label{sec:AsBias}
The FE program \eqref{eq:fe} can be seem as a special case of \eqref{eq:mest} with $j = it$, $n=NT$, $\theta=(\beta, \alpha, \gamma)$ and $p = d_{\beta} + N + T$. However, the random sampling assumption of Section~\ref{sec:IIDexpansion} is not plausible for panel data. The presence of the unobserved effects  introduces heterogeneity in both dimensions, and  assuming independence is often too strong when one of the dimensions is time.
Some adjustments to the expansion in \eqref{SecondOrderExpansionIID} are therefore required.
We can still derive a second-order expansion $\widehat \theta - \theta_0  \approx  (NT)^{-1/2}  \psi_1 + (NT)^{-1}  \psi_2$ for $\widehat \theta=(\widehat \beta, \widehat \alpha, \widehat \gamma)$ and $\theta_0=(\beta_0, \alpha_0, \gamma_0)$,
but the expressions of $\psi_1$ and $ \psi_2$ need to be modified to account for
 heterogeneity and weak serial dependence. We provide these modifications in the appendix.

To characterize the asymptotic bias of the component $\widehat \beta$ of $\widehat \theta$, 
it is convenient to make $\ell_{it}(\beta,\alpha,\gamma)$ information-orthogonal between $\beta$ and
 the rest of the parameters. This can be achieved by the transformation\footnote{
 This  transformation corresponds to the reparameterization
 $\alpha_i^* = \alpha_i -  \kappa_i'  \beta$ 
   and $\gamma_t^* = \gamma_t  -  \rho_t'  \beta$.
 The log-likelihood with respect to these parameters is
  $ \ell_{it}(\beta,    \alpha_i^* +   \kappa_i'  \beta ,     \gamma_t^* +  \rho_t'  \beta ) =: \ell^*_{it}(\beta,   \alpha_i^*  , \gamma_t^* )$,
 which gives      \eqref{DefLstar} after
 renaming $(\alpha_i^* ,  \gamma_t^*)$
  as $(\alpha_i, \gamma_t)$ again.
 }
\begin{align}
    \ell^*_{it}(\beta,\alpha_i,\gamma_t)  &:=
      \ell_{it}(\beta,  \alpha_i   +  \kappa_i'  \beta , \gamma_t  +  \rho_t'  \beta) ,
    \label{DefLstar}
\end{align} 
where the $d_{\beta}$-vectors $\kappa_i$ and $\rho_t$ are a solution to the system of equations\footnote{
   The solution for $\kappa_i$ and $\rho_t$ may not be unique. 
   For example, if $\ell_{it}(\beta,\alpha_i,\gamma_t)  =\ell_{it}(\beta,\alpha_i + \gamma_t)  $,
   then the system does not uniquely determine $\kappa_i$ and $\rho_t$, but only $\kappa_i +  \rho_t$
   for all $i,t$. However, only $\kappa_i +  \rho_t$ will appear
   in our formulas for asymptotic bias and variance of $\widehat \beta$ in that case, so the non-uniqueness
   of $\kappa_i$ and $\rho_t$ is not important.
   }
\begin{align}
     \mathbb{E}_{T}  \left[ \overline \ell_{it}^{\, \beta \alpha_i}  
                      +  \kappa_i  \,  \overline  \ell_{it}^{\, \alpha_i \alpha_i}  
                      +  \rho_t \,  \overline  \ell_{it}^{\, \alpha_i \gamma_t}  \right] &= 0 ,
          \qquad i=1,\ldots,N ,\notag
     \\          
     \mathbb{E}_{N}  \left[ \overline \ell_{it}^{\, \beta \gamma_t}  
                      +  \kappa_i  \,  \overline  \ell_{it}^{\, \gamma_t \alpha_i}  
                      +  \rho_t \,  \overline  \ell_{it}^{\, \gamma_t \gamma_t}  \right] &= 0 ,
          \qquad t=1,\ldots,T .
     \label{eq:orth}     
\end{align}
Here, as in Section~\ref{sec:IIDexpansion},
we drop the arguments of the partial derivatives of $\ell_{it}$ when they are evaluated at the true 
values $(\beta_0, \alpha_{0i}, \gamma_{0t})$,
and analogously we drop the arguments of partial derivatives of $\ell_{it}^*$ 
when they are evaluated at the transformed true values 
$(\beta_0, \alpha_{0i} -  \kappa_i'  \beta_0, \gamma_{0t} -  \rho_t'  \beta_0)$.
Solving the program  \eqref{eq:fe} with $\ell^*_{it}(\beta,\alpha_i,\gamma_t)$ in place of $ \ell_{it}(\beta,\alpha_i,\gamma_t)$ does not change the solution for $\beta$.
The reparametrized program  cannot be used to compute the estimator because  $ \ell^*_{it}(\beta,\alpha_i,\gamma_t)  $
depends on $\theta_0$, but it is a convenient theoretical device to analyze the properties of $\widehat \beta$ because the Hessian $ \mathbb{E}_{NT} \overline \ell^{* \, \theta \theta}_{it} $
is  block-diagonal between $\beta$ and $(\alpha,\gamma)$. We  proceed by obtaining the bias from the  infeasible log-likelihood $\ell_{it}^*$, and then we can express it in terms of the feasible likelihood $\ell_{it}$ using the one-to-one relationship between  $\ell_{it}$ and $\ell_{it}^*$. For example, $\ell_{it}^{* \, \beta} = \ell_{it}^{\beta} +  \kappa_i  \ell_{it}^{\alpha_i} + \rho_t  \ell_{it}^{\gamma_i}$.

Under the orthogonal parametrization, the component of the influence function $  \psi_1  =  \allowbreak -  (  \mathbb{E}_{NT} \overline \ell^{* \, \theta \theta}_{it})^{-1} \allowbreak  \GNT  \ell^{*\, \theta}_{it}  $ corresponding to $\beta$ simplifies to
 $\psi_{1,\beta} = - (  \mathbb{E}_{NT} \overline \ell^{* \, \beta \beta}_{it}  )^{-1}  \GNT  \ell^{*\, \beta}_{it}. $
Under standard conditions for MLE, as $N,T \to \infty$, 
$$
\psi_{1,\beta} \to_d {\cal N}(0, H^{-1} ), \qquad H =  - \plim_{N,T \rightarrow \infty} \mathbb{E}_{NT} \overline \ell^{*\, \beta \beta}_{it},
$$ 
where we use that $\psi_{1,\beta}$ is a martingale difference over $t$, and the conditional information equality. As in the cross-sectional case,  $\psi_{1,\beta} $ has zero mean and  determines the  asymptotic variance.

The component of the second term $\psi_2$ corresponding to $\beta$ also  simplifies with the reparametrization. After some calculations that are detailed in the appendix,  
\begin{eqnarray*}
  \psi_{2,\beta} &:=  \underset{ =  \psi^B_{2,\beta}}{\underbrace{- \left( \mathbb{E}_{NT} \overline \ell^{*\, \beta \beta}_{it} \right)^{-1} \sum_{i=1}^N   
         \left[ - 
          \frac{    \GT   \ell^{* \, \beta \alpha_i}_{it} \GT  \ell^{\alpha_i}_{it}    }
                   {   \ET \overline \ell^{\, \alpha_i  \alpha_i}_{it}     }        
                      + 
                      \frac{   \ET \overline \ell^{* \, \beta \alpha_i \alpha_i}_{it} 
                      \left( \GT  \ell^{\alpha_i}_{it} \right)^2 }
                      {2 \left( \ET \overline \ell^{\, \alpha_i  \alpha_i}_{it} \right)^2 } \right] }}
           \\ \nonumber & \;      \; \; \;
                \underset{ =  \psi^D_{2,\beta}}{\underbrace{- \left( \mathbb{E}_{NT} \overline \ell^{*\, \beta \beta}_{it} \right)^{-1}  \sum_{t=1}^T
         \left[
         -  \frac{  \GN \ell^{* \, \beta \gamma_t}_{it} 
                     \GN  \ell^{\gamma_t}_{it}     }
                   {   \EN \overline \ell^{\, \gamma_t  \gamma_t}_{it}     }      
                      + 
                      \frac{   \EN  \overline \ell^{* \, \beta \gamma_t \gamma_t}_{it} 
                      \left(  \GN  \ell^{\gamma_t}_{it} \right)^2 }
                      {2 \left( \EN \overline \ell^{\, \gamma_t  \gamma_t}_{it} \right)^2 }     
                      \right]}}.
\end{eqnarray*}
This second-order term is the primary source of bias. By the law of large numbers for heterogenous weakly dependent sequences as $N,T \to \infty$, 
\begin{equation}\label{eq:B}
  \frac{\psi^B_{2,\beta}}{N} \to_P   H^{-1}  \plim_{N,T \rightarrow \infty}
       \EN
        \left\{
         \frac{ -  \ET \sum_{s=t}^T   \Ep \left( \ell^{\alpha_i}_{it}  
               \ell^{*\, \beta \alpha_i}_{is}  
                 \right)
                 -  \ET   \overline  \ell^{*\, \beta \alpha_i \alpha_i}_{it} /2                
                 }
                              {  \mathbb{E}_T \, \overline \ell^{\, \alpha_i \alpha_i}_{it} }   
          \right\}        =: B,      
\end{equation}
and 
\begin{equation}\label{eq:D}
  \frac{\psi^D_{2,\beta}}{T} \to_P    H^{-1}  \plim_{N,T \rightarrow \infty}
       \ET
        \left\{
         \frac{   -  \EN \Ep \left( \ell^{\gamma_t}_{it}  
               \ell^{*\, \beta \gamma_t}_{it}  
                 \right)
                 -  \EN  \overline  \ell^{*\, \beta \gamma_t \gamma_t}_{it} /2                
                 }
                              {  \mathbb{E}_N \, \overline \ell^{\, \gamma_t \gamma_t}_{it} }   
          \right\}        =: D.      
\end{equation}
The expressions for $B$
and $D$ are almost symmetric with respect to the indices $i$ and $t$, 
but there is no double sum in $D$ because we are assuming independence of observations across $i$,
while we allow for predetermined covariates over $t$. The expressions become symmetric  when the covariates are strictly exogenous because the terms of the double sum $ \sum_{s=t+1}^T$ in $B$ drop out.  Hence, the order of the bias is
$$
\frac{\psi_{2,\beta}}{NT} = \frac{\psi^B_{2,\beta} + \psi^D_{2,\beta}}{NT} \sim \frac{B}{T} + \frac{D }{N}  \sim   \frac{N+T}{NT},
$$
which corresponds to the prediction from \eqref{BiasOrder} since $n=NT$ and $p = d_{\beta}+N+T \sim N+T$.

The bias term  $B/T$ comes from the individual fixed effects since there are $T$ observations that are informative about each of them. Symmetrically, the bias term  $D/N$ comes from the time fixed effects since there are $N$ observations that are informative about each of them. Accordingly, the term $D/N$ drops out in models with only individual effects or becomes negligible relative to $B/T$ when $N \gg T$. \citet*{Hahn:2002p717} 
and \citet*{AlvarezArellano2003} characterized $B$ in  dynamic linear panel models with individual effects. For nonlinear models, \citet*{Hahn:2004p882}  and  \citet*{HahnKuersteiner2011}  derived $B$ in the static and dynamic case, respectively. \cite*{FernandezValWeidner2016} characterized $B$ and $D$ in static and dynamic nonlinear models with individual and time effects.

The expressions of $B$ and $D$ can be further characterized  in specific models. 
\paragraph*{Example (i)} In the linear model \eqref{eq:linear},
$$
B = H^{-1} \plim_{N,T \to \infty} - \ENT \left\{ \sum_{s=t+1}^T \varepsilon_{it} \tilde x_{is}\right\}, \quad
 D = 0,  \quad
   H = \plim_{N,T \to \infty}  \ENT \left\{\tilde x_{it} \tilde x_{it}'\right\},
$$
where $\tilde x_{it}$ is the residual of the linear projection of $x_{it}$ on the space spanned by the individual and time effects
(i.e. the two-way demeaned $x_{it}$ in the linear model that corresponds to the orthogonal transformation). When the covariates are strictly exogenous, $B=0$, i.e., there is no incidental parameter problem.  
 \citet*{Hahn:2002p717} obtained this large $T$ bias expansion of the \citet*{Nickell81} bias  for dynamic linear panel models with only individual effects. 
\citet*{hahn2006reducing} showed that the bias expression carries over to dynamic linear panel models with individual and time effects because $D=0$.

\paragraph*{Example (ii)} In the  binary response model \eqref{eq:probit} with standard normal link $F_{\varepsilon}=\Phi$ or probit model,  when the covariates are strictly exogenous 
 $$
B = H^{-1} \ENT \left\{ \frac{\omega_{it} \tilde x_{it} \tilde x_{it}'}{\ET \omega_{it}}\right\} \beta_0, \ D = H^{-1} \ENT \left\{ \frac{\omega_{it} \tilde x_{it} \tilde x_{it}'}{\EN \omega_{it}} \right\} \beta_0, \ H=   \ENT \left\{\omega_{it} \tilde x_{it} \tilde x_{it}'\right\}, 
$$
where $\tilde x_{it}$ is the residual of the linear projection of $x_{it}$ on the space spanned by the individual and time effects under a metric weighted by $\omega_{it}$, $\omega_{it} = \phi_{it}^2/[\Phi_{it} (1 - \Phi_{it})]$, and $\phi_{it}$ and $\Phi_{it}$ are the standard normal PDF and CDF evaluated at $x_{it}'\beta_0 + \alpha_{0i} + \gamma_{0t}$. 
In this case there is bias, which is a positive definite matrix weighted average of $\beta_0$.  \citet*{FernandezVal:2009p3313} derived this result in models with individual effects and \cite*{FernandezValWeidner2016}  in models with individual and time effects.

\paragraph*{Example (iii)}  In the Poisson model \eqref{eq:poisson},
$$
B = H^{-1} \plim_{N,T \to \infty} - \ENT \left\{\frac{\sum_{s=t+1}^T (y_{it} - \lambda_{it}) \lambda_{is} \tilde x_{is}}{\EN \lambda_{it}}\right\}, \ D =0, \  H = \plim_{N,T \to \infty}  \ENT \left\{\lambda_{it} \tilde x_{it} \tilde x_{it}'\right\},
$$
where $\tilde x_{it}$ is the residual of the linear projection of $x_{it}$ on the space spanned by the individual and time effects under a metric weighted by $\lambda_{it}$. When the covariates are strictly exogenous, $B=0$,  i.e., there is no incidental parameter problem.  This is a well-known result for models with only individual effects
(e.g., \citealt{Palmgren1981}),\footnote{
 \citealp*{HausmanHallGriliches1984} use the conditional likelihood approach to eliminate the incidental
 parameters, which for the Poisson models turns out to be equivalent to fixed effect MLE,
 see \citet*{BlundellGriffithWindmeijer1999,BlundellGriffithWindmeijer2002} and
  \citet*{Lancaster:2002p875}.
}
which was extended to models with individual and time effect in \cite*{FernandezValWeidner2016}.

Finally, combining the properties of $\psi_{1,\beta}$  and $\psi_{2,\beta}$, we conclude that as $N,T \to \infty$ such that $N/T \to \kappa$, $0 < \kappa < \infty$, 
\begin{equation}\label{eq:adfe}
\widehat \beta - \beta_0 \overset{a}{\sim} {\cal N}\left(   \frac B T + \frac D N, \frac{H^{-1}} {NT} \right). 
\end{equation} 
This asymptotic approximation prescribes  that the fixed effects estimator can have significant  bias relative to its standard deviation. Moreover, by the argument given in Section \ref{subsec:inf}, confidence intervals constructed around the fixed effects estimator can severely undercover the components of $\beta_0$ even in large samples. \cite*{FernandezValWeidner2016} showed that these prescriptions provide a good approximations to the  behaviour of the fixed effects estimator for small sample sizes through analytical and simulation examples.

\subsection{Bias Corrections}\label{subsec:bc}
The goal of the bias corrections is to remove the bias from the asymptotic distribution \eqref{eq:adfe}, ideally without increasing the  variance. In other words, we want to find a bias corrected estimator   $\widehat \beta_{\rm BC}$ such that as $N,T \to \infty$,
$$
\widehat \beta_{\rm BC} - \beta_0 \overset{a}{\sim} {\cal N}\left(  0, \frac{H^{-1}} {NT} \right). 
$$
The corrected estimator therefore has small bias relative to its dispersion and the same variance as the FE estimator in large samples. Moreover, the confidence intervals constructed around the bias corrected estimator should have coverage close to their nominal level. We describe analytical and resampling methods to carry out the bias corrections. These methods  rely on the asymptotic distribution \eqref{eq:adfe}, together with consistent estimators of the bias.

\subsubsection{Analytical Bias Correction} 
The analytically bias corrected (ABC) estimator is
$$
\widehat \beta_{\rm ABC}  = \widehat \beta - \frac{\widehat B}{T} - \frac{\widehat D}{N},
$$
where $\widehat B$ and $\widehat D$ are consistent estimators of $B$ and $D$, i.e. $\widehat B \to_P B $ and $\widehat D \to_P D $ as $N,T \to \infty$. In this case, as $N,T \to \infty$ such that $N/T \to \kappa$, $0 < \kappa < \infty,$
$$
\widehat \beta_{\rm ABC} - \beta_0 \approx \frac{\psi_{1,\beta}}{\sqrt{NT}} + \frac{1}{T} \left( \frac{\psi^B_{2,\beta}}{N} - \widehat B\right) +\frac{1}{N} \left( \frac{\psi^D_{2,\beta}}{T} - \widehat D\right) \approx \frac{\psi_{1,\beta}}{\sqrt{NT}},
$$
since $\psi^B_{2,\beta}/N - \widehat B \to_P 0$ and $\psi^D_{2,\beta}/T - \widehat D \to_P 0$. Hence, 
$$
\widehat \beta_{\rm ABC} - \beta_0 \overset{a}{\sim} {\cal N}\left(  0, \frac{H^{-1}} {NT} \right).
$$

The estimators  $\widehat B$ and $\widehat D$ are constructed from the analytical expressions of $B$ and $D$ given in \eqref{eq:B} and \eqref{eq:D}. 
Let  $\widehat \ell_{it} = \ell_{it}(\widehat \beta,\widehat \alpha_i, \widehat \gamma_t)$
 denote the log-likelihoods evaluated at the FE estimators, and define their  derivatives evaluated at the FE estimators analogously. Let
$\widehat \ell^*_{it}(\beta,\alpha_i,\gamma_t)  :=
      \ell_{it}(\beta,  \alpha_i   +  \widehat \kappa_i'  \beta , \gamma_t  +  \widehat \rho_t'  \beta)$,
with       $\widehat \kappa_i$ and $\widehat \rho_t$ defined analogously to $\kappa_i$ and $\rho_t$ in \eqref{eq:orth},
but using the fixed effect estimates instead of the true value of the parameters.
All derivatives of       $\widehat \ell^*_{it}$ are evaluated
at $(\widehat \beta,\widehat \alpha_i -  \widehat \kappa_i'  \widehat \beta, 
  \widehat \gamma_t -  \widehat \rho_t'  \widehat \beta)$ in the following.
Then, the plug-in estimators of $B$ and $D$ are
\begin{equation*}
 \widehat B =   \widehat H^{-1}  
       \ENT
        \left\{
         \frac{ \sum_{s=t}^{t+M \wedge T}    \widehat \ell^{\alpha_i}_{it}  
               \widehat \ell^{*\, \beta \alpha_i}_{is}  
                 -   \widehat  \ell^{*\, \beta \alpha_i \alpha_i}_{it} /2                
                 }
                              {  \mathbb{E}_T \, \widehat \ell^{\, \alpha_i \alpha_i}_{it} }   
          \right\} ,      \\
\end{equation*}
and 
\begin{equation*}
  \widehat D =     \widehat H^{-1} 
       \ENT
        \left\{
         \frac{   \widehat \ell^{\gamma_t}_{it}  
               \widehat \ell^{*\, \beta \gamma_t}_{it}  
                 -   \widehat  \ell^{*\, \beta \gamma_t \gamma_t}_{it} /2                
                 }
                              {  \mathbb{E}_N \, \widehat \ell^{\, \gamma_t \gamma_t}_{it} }   
          \right\},      \\
\end{equation*}
where $\widehat H = -  \mathbb{E}_{NT} \widehat \ell^{*\, \beta \beta}_{it}$, and $M$ is a trimming parameter such that $M/T \to 0$ and $M\to\infty$ as $T \to \infty$ \citep*{HK2007}.   We can set $M=0$ when the covariates are strictly exogenous. These estimators use  $\widehat \beta$, i.e. $\widehat B = \widehat B(\widehat \beta)$ and $\widehat D = \widehat D(\widehat \beta)$.  It is possible to iterate the correction by  (i) estimating $B$ and $D$ using $\widehat \beta_{\rm ABC}$ to obtain $\widehat B_2 = \widehat B(\widehat \beta_{\rm ABC})$, $\widehat D_2 = \widehat D(\widehat \beta_{\rm ABC})$ and $\widehat \beta_{\rm ABC2} = \widehat \beta - \widehat B_2/T - \widehat D_2/N$; (ii) estimating $B$ and $D$ using $\widehat \beta_{\rm ABC2}$  to obtain $\widehat B_3 = \widehat B(\widehat \beta_{\rm ABC2})$, $\widehat D_3 = \widehat D(\widehat \beta_{\rm ABC2})$ and $\widehat \beta_{\rm ABC3} = \widehat \beta - \widehat B_3/T - \widehat D_3/N$; and so on.  The iteration does not affect the asymptotic distribution of the correction but might improve its small sample properties.

\citet*{HahnKuersteiner2011}  developed the ABC for general dynamic nonlinear models with unobserved individual effects, building on the analysis of   \citet*{Hahn:2002p717} for dynamic linear models and \citet*{Hahn:2004p882} for static nonlinear models. \cite*{FernandezValWeidner2016} extended the ABC to models with unobserved individual and time effects.

\subsubsection{Leave-One-Out Jackknife Bias Correction} 
\label{sec:LeaveOneOut}
This correction is based on the Jackknife method introduced by \citet*{quenouille1956notes} and \citet*{tukey58} for cross-sectional data. We start by giving some intuition on how this method works using the example of Section \ref{sec:bias}.\footnote{We refer to \cite{ShaoTu1995} for a rigourous analysis of the properties of the Jackknife.} Let $\widehat \theta_{(-j)}$ denote the estimator of $\theta$ that leaves out the $j^{th}$ observation.  The leave-one-out  bias corrected estimator is
$$
\widehat \theta_{\rm JC} = n  \widehat \theta - (n-1) \bar{\theta}_{n-1}, \ \ \bar{\theta}_{n-1} = \frac{1}{n} \sum_{j=1}^n \widehat \theta_{(-j)}. 
$$
To understand how the  correction works, assume the second-order expansion for the bias $\Ep \widehat \theta - \theta_0 = B_1/n + B_2/n^{2} + o(n^{-2})$. Then, under identical distribution $\Ep \widehat \theta_{(-j)} - \theta_0 = B_1/(n-1) + B_2/(n-1)^{2} + o(n^{-2})$ for all $j$, so that
$$
\Ep \widehat \theta_{\rm JC} - \theta_0 = B_1 + \frac{B_2}{n} - B_1 - \frac{B_2}{n-1} + o(n^{-1}) = o(n^{-1}).
$$
In other words $(n-1)(\bar{\theta}_{n-1} - \widehat \theta)$ is an estimator of the first-order bias since
$
(n-1)\Ep(\bar{\theta}_{n-1} - \widehat \theta) = B_1/n + o(n^{-1}).
$

 \citet*{Hahn:2004p882} introduced the Jackknife to panel models with individual effects, and \cite*{FernandezValWeidner2016} extended it to  panel models with individual and time effects. To describe how to apply the Jackknife to panel data, it is convenient to introduce some notation. Let ${\bf N} =  \{ 1,\ldots, N\}$ and ${\bf T} =  \{ 1,\ldots, T\}$ be the sets of indexes for the two dimensions of the panel. For the subsets of indexes $A \subseteq {\bf N}$ and $C \subseteq {\bf T}$, let $\widehat \beta_{A,C}$ denote the FE estimator of $\beta$ in the subpanel with indexes $(i,t) \in A \times C = \{(i,t): i \in A, t \in C\}$, that is
\begin{align*}
\widehat \beta_{A,C} =  \argmax_{\beta \in \mathbb{R}^{d_{\beta}}} \;
   \max_{\alpha \in \mathbb{R}^{|A|}}  \;
    \max_{\gamma \in \mathbb{R}^{|C|}} 
\; \sum_{i \in A,t \in C}   \;  \ell_{it}(\beta,\alpha_i,\gamma_t),
\end{align*}
where $|A|$ denotes the cardinality of the set $A$.  With this notation the FE estimator of $\beta$ is $\widehat \beta = \widehat \beta_{{\bf N},{\bf T}}$. Define the average leave-one-out estimators in each dimension as
\begin{align*}
       \overline{\beta}_{N-1,T}
       &= \frac 1 N \sum_{i=1}^N 
           \,  \widehat \beta_{{\bf N} \setminus \{i\}, {\bf T}}  ,
       &    
     \overline{\beta}_{N,T-1}
       &= \frac 1 T \sum_{t=1}^T 
           \,  \widehat \beta_{{\bf N}, {\bf T} \setminus \{t\}} .
\end{align*}
The panel Jackknife estimator for models with individual effects is  $T   \widehat \beta_{{\bf N},{\bf T}} -  (T-1)\overline \beta_{N,T-1}$. The correction $(T-1)(\overline \beta_{N,T-1} -   \widehat \beta_{{\bf N},{\bf T}})$ removes the bias term  $B/T$, analogously to the cross-sectional case discussed above.
A panel jackknife bias corrected (JBC) estimator
that removes both bias terms can be formed as
$$
\widehat \beta_{\rm JBC}  =  (N+T-1) \widehat \beta_{{\bf N},{\bf T}} 
- (N-1)   \overline \beta_{N-1,T} - (T-1)   \overline \beta_{N,T-1},
$$
This correction removes the bias in large samples because $(N-1)\Ep(\overline{\beta}_{N-1,T} - \widehat \beta_{{\bf N},{\bf T}} ) \approx D/N$ and $(T-1)\Ep(\overline{\beta}_{N,T-1} - \widehat \beta_{{\bf N},{\bf T}} ) \approx B/T$. Moreover, it can be shown that under suitable conditions
$$
\widehat \beta_{\rm JBC}  - \beta_0 \overset{a}{\sim} {\cal N}\left(  0, \frac{H^{-1}} {NT} \right).
$$
An important limitation in the application of the leave-one-out Jackknife to panel data is that it requires independence in both dimensions of all the variables. This is a very restrictive condition for panel data, specially when one of the dimensions is time,
as it rules out lagged-dependent variables or serially correlated variables as covariates.

\subsubsection{Split-Sample Jackknife Bias Correction} This correction is based on the split-sample Jackknife method introduced by \citet*{quenouille1949approximate} for time series data. We again start by giving an intuitive description of this method in the context of the simple model of Section \ref{sec:bias}. Assume that the sample size $n$ is even and split the sample in two halves: $\{z_j : 1 \leq j \leq n/2\}$ and $\{z_j : n/2 + 1 \leq j \leq n\}$.  Let $\widehat \theta^1$ and $\widehat \theta^2$ denote the estimator of $\theta$  in each of the half-samples.  The split-sample jackknife bias  corrected estimator is
$$
 \widehat \theta_{\rm SC} = 2  \widehat \theta -  \widetilde{\theta}_{1/2} = \widehat \theta -  (\widetilde{\theta}_{1/2} - \widehat \theta), \qquad  \widetilde{\theta}_{1/2} = (\widehat \theta^1 + \widehat \theta^2)/2.
$$
Assume the first-order expansion for the bias $\Ep \widehat \theta - \theta_0 = B/n + o(n^{-1})$. The correction works asymptotically  because under stationarity $\Ep \widehat \theta^1 = \Ep \widehat \theta^2  = \theta_0  +  2B/n + o(n^{-1}) $, so that
$$
\Ep \widehat \theta_{\rm SC} - \theta_0 =  \frac{2B}{n} - \frac{2B}{n} + o(n^{-1}) = o(n^{-1}).
$$
In other words $\widetilde{\theta}_{1/2} - \widehat \theta$ is an estimator of the bias since
$
\Ep(\widetilde{\theta}_{1/2} - \widehat \theta) = B/n + o(n^{-1}).
$

To define the panel version of this Jackknife correction we introduce
\begin{align*}
  \widetilde{\beta}_{N/2,T}
       &= \frac 1 2 
       \left[  \widehat \beta_{\{i \leq \lceil N/2 \rceil \}, {\bf T}}
            +   \widehat \beta_{ \{i  \geq \lfloor N/2 + 1 \rfloor  \}, {\bf T}}
       \right] ,
       &    
\widetilde{\beta}_{N,T/2}
       &= \frac 1 2 
       \left[  \widehat \beta_{{\bf N}, \{t \leq \lceil T/2 \rceil \}}
            +   \widehat \beta_{{\bf N}, \{t  \geq \lfloor T/2 + 1\rfloor  \}}
       \right],
\end{align*}
where $\lceil . \rceil$  and  $\lfloor . \rfloor$ are the ceiling and floor functions,
and we use the same notation as in the previous section.
\citet*{DhaeneJochmans2015} introduced the split-sample jackknife to panel models with individual effects. In this case the split-sample bias corrected (SBC) estimator is  $2 \widehat \beta_{{\bf N},{\bf T}} -  \widetilde \beta_{N,T/2} $. The correction $\widetilde \beta_{N,T/2} - \widehat \beta_{{\bf N},{\bf T}}$ removes the bias term $B/T$, analogously to the cross-sectional case. \cite*{FernandezValWeidner2016}  extended the SBC estimator
to panel models with individual and time effects, in which case the correct linear combination is
$$
\widehat \beta_{\rm SBC}  =  3 \widehat \beta_{{\bf N},{\bf T}} -   \widetilde \beta_{N/2,T} -    \widetilde \beta_{N,T/2}.
$$
This correction removes both bias terms in large samples because $\Ep(\widetilde{\beta}_{N/2,T} - \widehat \beta_{{\bf N},{\bf T}} ) \approx D/N$ and $\Ep(\widetilde{\beta}_{N,T/2} - \widehat \beta_{{\bf N},{\bf T}} ) \approx B/T$. The intuition here is simple. The estimator $\widetilde{\beta}_{N,T/2}$ has double the bias  than $\widehat \beta_{{\bf N},{\bf T}}$ coming from the estimation of the individual effects, because there are only half of observations, $T/2,$ informative about each of them. However, $\widetilde{\beta}_{N,T/2}$ has the same bias as $\widehat \beta_{{\bf N},{\bf T}}$ coming from the estimation of the time effects, because there are the same number of observations, $N$, informative about each of them. A similar argument shows that the bias of $ \widetilde \beta_{N/2,T} $ is $2D/N + B/T$. Under suitable conditions,
$$
\widehat \beta_{\rm SBC}  - \beta_0 \overset{a}{\sim} {\cal N}\left(  0, \frac{H^{-1}} {NT} \right).
$$
\citet*{quenouille1956notes} already noted that Jackknife correction can be extended to higher order,
and  \citet*{DhaeneJochmans2015} discussed high-order Jackknife corrections, 
that is, elimination not only of the leading bias of order $T^{-1}$, but also of the next order bias term of order $T^{-2}$,
which can again be achieved by an appropriate linear combination of split-panel fixed-effect estimates.
A formal discussion of those higher-order bias terms   is given in
\citet*{BunKiviet2003} for dynamic linear models panel models
and in \citet*{SunDhaene2017} for non-linear models.

\subsubsection{Hybrid Jackknife Bias Correction}  When the data are independent across one of the dimensions of the panel, it is not clear how to carry out the sample split in this dimension. One possibility is to make multiple splits and average the resulting corrected estimators for each split. Another possibility is a hybrid method that uses leave-one-out  along the independent dimension and split-sample along the weakly dependent dimension. For example, if we assume independence over the cross-sectional dimension, then a hybrid jacknife estimator is
$$
     \widehat {\beta}_{\rm HBC} = (N+1)\widehat{\beta}_{{\bf N},{\bf T}} - (N-1)\overline{\beta}_{N-1,T} - \widetilde{\beta}_{N,T/2}.
$$
This correction removes the bias in large samples because $(N-1)\Ep(\overline{\beta}_{N-1,T} - \widehat \beta_{{\bf N},{\bf T}} ) \approx D/N$ and $\Ep(\widetilde{\beta}_{N,T/2} - \widehat \beta_{{\bf N},{\bf T}} ) \approx B/T$. Again, under suitable conditions
$$
\widehat \beta_{\rm HBC}  - \beta_0 \overset{a}{\sim} {\cal N}\left(  0, \frac{H^{-1}} {NT} \right).
$$
We refer to \cite*{Stata2015} for other examples of hybrid corrections.

\subsubsection{Comparison}  All the corrections produce estimators with the same asymptotic distribution, but they rely on different sampling conditions and might involve the choice of tuning parameters. Compared to the JBC; ABC, SBC and HBC work under more general sampling conditions because they do not require independence of the data along both dimensions. However, SBC and HBC usually increase dispersion  in small samples because they estimate the parameters in smaller panels than JBC. Compared to ABC, SBC and HBC work under less general sampling conditions because they rely on homogeneity in both dimensions in order for the bias to be the same in all the subpanels. ABC does not use homogeneity, but requires coding the estimators of the bias, which involves a delicate choice for the trimming parameter $M$ when the covariates are not  strictly exogenous. In practice, we recommend to always carry out a sensitivity analysis reporting results from both jackknife and analytical corrections and to try several values of $M$ in the analytical correction starting with $M=0,1,2,3,4$. 

Some of the differences can be appreciated in a simple cross sectional example where we can characterize the moments of all the estimators. Consider the normal model $z \sim \mathcal{N}(\mu,\sigma^2)$, $\sigma^2 > 0$. The MLE of $\sigma^2$ is the sample variance $\widehat \sigma^2 = n^{-1} \sum_{j=1}^n (z_j - \bar z)^2,$ where $\bar z = n^{-1} \sum_{j=1}^n z_j$. As it is well-known, $\widehat \sigma^2$ is biased with ${\rm Bias}(\widehat \sigma^2)  = - \sigma^2/n$ and $\mathrm{Var}(\widehat \sigma^2 ) = 2 \sigma^4 (n-1)/n^2$. The ABC is $\widetilde \sigma^2_{\rm ABC} = (n+1) \widehat \sigma^2/n$, which has ${\rm Bias}( \widetilde \sigma^2_{\rm ABC})= - \sigma^2/n^2$ and $\mathrm{Var}(\widetilde \sigma^2_{\rm ABC} ) =  (n+1)^2\mathrm{Var}(\widehat \sigma^2 )/n^2$. Iterating $k$ times yields  $\widetilde \sigma^2_{\rm ABCk} = (\sum_{r=0}^{k} n^{-r}) \widehat \sigma^2$, which has ${\rm Bias}(\widetilde \sigma^2_{\rm ABCk})= - \sigma^2/n^{k+1}$ and $\mathrm{Var}(\widetilde \sigma^2_{\rm ABCk} ) = (\sum_{r=0}^{k} n^{-r})^2 \mathrm{Var}(\widehat \sigma^2 )$. The JBC can be shown to be $\widetilde \sigma^2_{\rm JBC} = n \widehat \sigma^2/(n-1),$ the degrees-of-freedom corrected estimator, which has ${\rm Bias}( \widetilde \sigma^2_{\rm JBC})= 0$ and $\mathrm{Var}(\widetilde \sigma^2_{\rm JBC} ) = n^2 \mathrm{Var}(\widehat \sigma^2 )/(n-1)^2$. Let $\bar z_1$ and $\bar z_2$ be the sample means of $z$ in the first and second halves of the sample.  Simple algebra shows that the SBC is $\widetilde \sigma^2_{\rm SBC} = \widehat \sigma^2 + \bar z^2 - \bar z_1 \bar z_2,$  which has ${\rm Bias}( \widetilde \sigma^2_{\rm SBC})= 0$ and $\mathrm{Var}(\widetilde \sigma^2_{\rm SBC} ) = (n+2) \mathrm{Var}(\widehat \sigma^2 )/n$. A comparison of biases and variances reveals that in this case  ${\rm Bias}(\widehat \sigma^2)  > {\rm Bias}( \widetilde \sigma^2_{\rm ABC}) > {\rm Bias}( \widetilde \sigma^2_{\rm ABCk}) > {\rm Bias}( \widetilde \sigma^2_{\rm JBC}) = {\rm Bias}( \widetilde \sigma^2_{\rm SBC})$, while $\mathrm{Var}(\widehat \sigma^2 ) < \mathrm{Var}(\widetilde \sigma^2_{\rm ABC} ) < \mathrm{Var}(\widetilde \sigma^2_{\rm ABCk} ) < \mathrm{Var}(\widetilde \sigma^2_{\rm JBC} ) < \mathrm{Var}(\widetilde \sigma^2_{\rm SBC} )$.

\begin{table}[!hbt]
\centering
\caption{Probit Model with Strictly Exogenous Covariates ($N=664$, $T = 9$)}\label{tab:sim}
\begin{tabular}{l*{9}{c}}
\hline\hline
&\multicolumn{1}{c}{Bias}&\multicolumn{1}{c}{SD}&\multicolumn{1}{c}{RMSE}&\multicolumn{1}{c}{p; 95}&&\multicolumn{1}{c}{Bias}&\multicolumn{1}{c}{SD}&\multicolumn{1}{c}{RMSE}&\multicolumn{1}{c}{p; 95}\\ \hline
&\multicolumn{4}{c}{FE} & & \multicolumn{4}{c}{ABC}\\
kids0\_2	&14.9	&9.9	        &17.9		&0.58	&&1.1	&8.6   	&8.6		&0.95\\
kids3\_5	&14.5	&14.0	&20.1		&0.80	&&0.8	&12.2	&12.2	&0.97\\
kids6\_17	&14.6	&35.8	&38.6		&0.92	&&1.3	&31.6	&31.5	&0.97\\
&\multicolumn{4}{c}{SBC (1 partition)} & & \multicolumn{4}{c}{SBC (50 partitions)}\\
kids0\_2			&-0.5	&11.5	&11.5		&0.84	&&-0.9	&11.9	&12.0		&0.84\\
kids3\_5			&-1.0  	&18.6	&18.6		&0.82	&&-1.7	&19.5	&19.5		&0.82\\
kids6\_17			&-2.1	&49.9	&49.9		&0.81	&&-4.2	&50.5	&50.7		&0.80\\
&\multicolumn{4}{c}{HBC} & & \multicolumn{4}{c}{JBC}\\
kids0\_2			&-1.1  	&11.2	&11.3		&0.88	&&-4.3	&7.9  	&9.0		&0.93\\
kids3\_5			&-0.8  	&17.9	&17.9		&0.84	&&-4.5	&12.2	&13.0	&0.95\\
kids6\_17			&-2.2	&48.5	&48.5		&0.82	&&-6.3	&31.0	&31.6	&0.97\\
\hline\hline
\multicolumn{10}{l}{\footnotesize{Notes: 500 simulations calibrated to the PSID 1980--1988. }} \\
\multicolumn{10}{l}{\footnotesize{Bias, SD and RMSE are in percentage of true value of the parameter. }} \\
\multicolumn{10}{l}{\footnotesize{Calculations in {\tt Stata} with the command {\tt probitfe} \citep{Stata2015}.}}
\end{tabular}
\end{table}

Table \ref{tab:sim} reports the results of a numerical simulation for a  probit model with strictly exogenous covariates and unobserved individual and time effects. The design is calibrated to the female labor force participation (LFP) application in \citet*{FernandezVal:2009p3313}.  Thus, we draw 500 panels of size $N=664$ and $T=9$, with 
$$
y_{it} = 1(\beta'x_{it} + \alpha_i + \gamma_t \geq \varepsilon_{it}), \ \ i = 1,\ldots, N, \ \ t = 1, \ldots, T, 
$$
where $x_{it}$ includes three fertility variables (the numbers of children aged 0-2, 3-5, and 6-17), the
logarithm of the husband's earnings in 1995 thousands of dollars, and a quadratic
function of age in years divided by 10, whose values are taken for the PSID 1980--1988;  and $\varepsilon_{it}$ are independent draws from the standard normal distribution.\footnote{The original PSID sample includes $1,461$ women, but only $664$ of them have variation in the LFP variable over the years of the panel.}   The parameters $(\beta,  \alpha_i, \gamma_t)$ are calibrated to the FE probit estimates in the PSID 1980--1988 with the observed LFP as the dependent variable. The table reports biases, standard deviations, root mean square errors,  and empirical  coverage probabilities of confidence intervals with nominal level of 95\% for the FE, ABC, SBC, HBC, and JBC estimators of the coefficients of the three fertility variables. We consider two versions of the SBC, one based on one partition over the cross sectional dimension following the ordering in the data set, and the other based on the average over 50 random partitions. All the results, except for the coverage probabilities, are in percentage of the true value of the parameter.  We find that the ABC and JBC drastically reduce  bias, dispersion and rmse, and have coverage probabilities close to their nominal level. The evidence for the SBC and HBC is more mixed. While  always reducing bias, they  increase dispersion resulting in higher rmse and lower coverage than the FE for the kids6\_17 fertility variable.  Increasing the number of partitions in the SBC does not improve the finite sample performance of this correction.

\subsubsection{Other  methods} 
The methods discussed so far correct the estimator. The same methods can also be applied to correct the first order conditions of the FE estimator of $\beta$. Namely, let $L(\beta) = \max_{(\alpha, \gamma) \in \mathbb{R}^{N+T}}  \,  \mathbb{E}_{NT}  \ell_{it}(\beta, \alpha_i,\gamma_t)$ be the profile objective function of \eqref{eq:fe}, where all the fixed effects have been concentrated out. Then, a second-order expansion similar to   \eqref{eq:adfe} yields
\begin{equation} 
   \frac{d L(\beta_0)} {d \beta} 
    \overset{a}{\sim} 
    {\cal N}\left(  - \frac {b} T - \frac {d} N, \, \frac{H} {NT} \right). 
\end{equation} 
where $b=HB$ and $d=HD$. This expression was derived in \citet*{BesterHansen2009} for models with individual effects and \citet*{Sun2016}
and \citet*{JochmansOtsu2016} for models with individual and time effects. Thus, the score of  $L(\beta)$ at the true parameter value is not centered at zero, which is the source of the bias in $\widehat \beta$. 
A ``profile-score bias corrected estimator'' is the solution to
$$d L(\widehat \beta_{\rm PSBC}) / d \beta + \widehat b/T + \widehat d/N= 0,$$ where $\widehat b= \widehat b(\widehat \beta)$
and $\widehat d =  \widehat d(\widehat \beta)$ are consistent estimators for $b$ and $d$, respectively. An alternative score correction that estimates simultaneously the parameter and bias terms can be formed as the solution to
$$
d L(\widehat \beta_{\rm PSBC2}) / d \beta + \widehat b(\widehat \beta_{\rm PSBC2})/T +  \widehat d(\widehat \beta_{\rm PSBC2})/N= 0.
$$
The analytical properties of this corrected profile-score are model specific, since they depend
on the functional form of $ \widehat b(\beta) $ and $ \widehat d(\beta) $. In particular, they may have multiple or no solution, even in the case of concave log-likelihood, where the solution for the previous 
equation for $\widehat \beta_{\rm PSBC}$ is unique.
\citet*{DhaeneJochmans2016likelihood}  discussed this issue in detail for linear autoregressive panel models,
and \cite*{dhaene2015profile} explored the behaviour of  these profile-score adjustments for other
panel data models with individual effects.  \citet*{DhaeneJochmans2015}  developed jackknife methods to correct the scores in models with individual effects.  

Finally, 
\citet*{gonccalves2015bootstrap} show that 
 bootstrap methods can correct for the Nickell bias in dynamic linear panel models,
and \citet*{KimSun2016} show how to use  the bootstrap to construct bias-corrected estimators
in non-linear panel models.

\subsection{Average Partial Effects}
The objects of interest in panel models are often \textit{ceteris paribus} or partial effects,  i.e. effects in the outcome of changing each covariate while holding the rest of covariates and unobserved effects fixed \citep*{Chamberlain1984}. In linear models the parameters correspond to partial effects. For example, in the model \eqref{eq:linear}, the components of $\beta$ measure the partial effects of each of the covariates. This effect is the same for all the individuals and time periods by  linearity. In nonlinear models the partial effects depend not only on the parameters but also on the covariates and unobserved effects. For example in the  nonlinear model for the conditional expectation $\Ep[y_{it} \mid x_{it},\alpha_i,\gamma_t] = m(x_{it}, \alpha_i, \gamma_t, \beta)$, the partial effect of changing the covariates from $x_{it}^0$ to $x_{it}^1$ is
$$
\delta_{it}(\alpha_{i},\gamma_{t}, \beta) = m(x^1_{it}, \alpha_{i}, \gamma_t,\beta) - m(x^0_{it}, \alpha_i, \gamma_t,\beta),
$$  
where $x^1_{it}$ and $x^0_{it}$ might depend on $x_{it}$. If $x_{it}$ is continuous and $x \mapsto m(x, \alpha_i, \gamma_t, \beta)$ a.s. differentiable,  the partial effect of a marginal change in the covariates is
$$
\delta_{it}(\alpha_i,\gamma_t, \beta) = \partial m(x, \alpha_i, \gamma_t)/\partial x \big|_{x = x_{it}}.
$$
In both cases the partial effects are heterogenous across $i$ and $t$.

\paragraph*{Example (ii)} In the binary response model, if the $k$th covariate $x_{k,it}$ is binary, then
the partial effect of changing $x_{k,it}$ from $0$ to $1$ on the probability of $y_{it}=1$ conditional on the rest of the covariates and unobserved effects is
$$
\delta_{it}(\alpha_i,\gamma_t, \beta) 
  = F_{\varepsilon}(\beta_k +
x_{it,-k}'\beta_{-k} +  \alpha_i + \gamma_t) -
F_{\varepsilon}(x_{it,-k}'\beta_{-k} + \alpha_i +  \gamma_t) ,
$$
where $\beta_k$ is the $k$th element of $\beta$, and $x_{it,-k}$ and $\beta_{-k}$ include all elements of $x_{it}$ and $\beta$ except for the $k$th element. 

\paragraph*{Example (iii)} In the count response example, if the $k$th covariate $x_{k,it}$ is continuous, then the  partial effect of a marginal change in $x_{k,it}$ on the average of $y_{it}$ conditional on the rest of the covariates and unobserved effects is
$$
\delta_{it}(\alpha_i,\gamma_t, \beta) 
= \beta_k  \exp(x_{it}'\beta +
\alpha_i + \gamma_t),
$$
where $\beta_k$ is the $k$th element of $\beta$.

One way to summarize the heterogeneity in the partial effects is to average them across the  individuals and  time periods in the panel. This yields the in-sample average partial effect (APE)
$$
\delta_{NT} = \ENT \delta_{it}(\alpha_i,\gamma_t, \beta). 
$$
Another possibility is to assume that there is a population of individuals and time periods from where the observed panel is drawn and consider the APE in this population. Under weak conditions this in-population APE corresponds to
$$
\delta = \plim_{N,T\to\infty} \delta_{NT}.
$$
Yet another possibility is to average the partial effects of the individuals in the population over the time periods in the sample
$$
\delta_T = \plim_{N\to\infty} \delta_{NT}.
$$
The choice of the relevant APE is application-specific.

The fixed effect estimator is the same for all the previous APEs. Thus, applying the plug-in principle,
$$
\widehat \delta_{NT} = \ENT \delta_{it}(\widehat \alpha_i, \widehat \gamma_t, \widehat \beta).
$$
However, the asymptotic distribution of $\widehat \delta_{NT} $ depends on the APE of interest. The asymptotic distribution around $\delta_{NT}$ is
$$
\widehat \delta_{NT} - \delta_{NT} \overset{a}{\sim} \mathcal{N}\left( \frac{E}{T} + \frac{F}{N},  \frac{\Sigma}{NT}\right),
$$
where the expressions for the bias terms $E$,  $F$,  and the variance $\Sigma$ 
are given in \cite*{FernandezValWeidner2016}.
The bias and variance here come from the estimation of the parameters and unobserved effects. 
The distribution around $\delta$ becomes
\begin{equation}\label{eq:ape}
\widehat \delta_{NT} - \delta  \overset{a}{\sim} \mathcal{N}\left(\frac{E}{T} + \frac{F}{N},  \frac{\Sigma + a_{NT} \Omega}{NT}\right),
\end{equation}
where the additional variance term,
$
\Omega = \plim_{N,T \to \infty} \left(\ENT [\delta_{it}(\alpha_i,\gamma_t,\beta) - \delta]\right)^2/a_{NT},
$
comes from the estimation of the population mean $\delta$ using the sample mean $\delta_{NT}$, and $a_{NT} = N \vee T$ because of the correlation of $\delta_{it}(\alpha_i,\gamma_t,\beta)$ with $\delta_{jt}(\alpha_j,\gamma_t,\beta)$ and $\delta_{is}(\alpha_i,\gamma_s,\beta)$ induced by the individual and time effects. Similarly, the distribution around $\delta_T$ is
$$
\widehat \delta_{NT} - \delta_{T} \overset{a}{\sim} \mathcal{N}\left( \frac{E}{T} + \frac{F}{N},  \frac{\Sigma + T \Omega_{T}}{NT} \right),
$$
where 
$
\Omega_{T} = \plim_{N \to \infty} \left(\ENT [\delta_{it} (\alpha_i,\gamma_t,\beta)- \delta_T]\right)^2/T.
$
Analogously to the parameters, the bias in the asymptotic distribution of the estimators of the APEs can be removed using analytical and jackknife corrections \citep*{FernandezValWeidner2016}.

The rate of convergence of  $\widehat \delta_{NT}$ differs depending on the APE of interest. Thus, $\delta_{NT}$ can be estimated at the rate $\sqrt{NT}$, the same rate as the parameter $\beta$, whereas $\delta$ can be estimated at the rate $\sqrt{NT/a_{NT}} = \sqrt{N} \wedge \sqrt{T}$ and $\delta_T$ at the rate $\sqrt{T}$. As a consequence of the slower rate of convergence, the  asymptotic distribution \eqref{eq:ape} simplifies to
\begin{equation}\label{eq:ape2}
\widehat \delta_{NT} - \delta \overset{a}{\sim} \mathcal{N}\left( 0,  \frac{a_{NT} \Omega}{ NT} \right),
\end{equation}
after dropping terms of lower order.\footnote{Similarly the asymptotic distribution around $\delta_T$ simplifies to $\widehat \delta_{NT}-\delta_{T} \overset{a}{\sim} \mathcal{N}(0 ,  \Omega_{T}/N)$.} In other words the estimation of the parameters  does not have first-order effect on the distribution around $\delta$. 
The reason is that the bias and standard deviation introduced by the parameter estimation are of lower order compared to  $\sqrt{NT/a_{NT}} $.  Despite this possible asymptotic simplification, we recommend using the higher-order distribution in \eqref{eq:ape} to perform inference on $\delta$, because it provides a more accurate approximation to the finite-sample distribution of $\widehat \delta_{NT}$ than the first-order distribution \eqref{eq:ape2}.

\subsection{Computation}\label{subsec:com}
The FE program \eqref{eq:fe} might look like computationally challenging due to the high dimension of the fixed effects. However, there are two aspects of the program that greatly facilitate the computation. First, the objective function is concave and smooth in the most commonly used cases such as the linear, Poisson, logit, probit, ordered probit, and tobit models. Second, the design matrix is sparse allowing the use of sparse algebra methods to speed up the computation. The SBC correction preserves the computational properties of the FE estimator as it only involves solving the program \eqref{eq:fe} in a small number of subsamples. The JBC and HBC can be computationally more intensive than the SBC when the cross-sectional dimension is large. The ABC also preserves the computational properties of the FE estimator, but requires to code the estimators of the bias that are model specific.

The bias corrections are available in the statistical software packages \verb"Stata" and \verb"R" for some models.  \cite*{Stata2015} provided two \verb"Stata" commands that implement the analytical and several jackknife corrections for estimators of parameters and APEs in logit and probit models with individual and time effects. These commands also report standard errors for the estimates of the parameter and APEs constructed using the asymptotic distribution.  \citet*{Sun2016stata} implemented in \verb"Stata"  the split-sample jackknife estimator and score corrections for parameters in linear, logit and probit models with individual effects. The command also has functionality for other single index user-supplied models and reports standard errors based on the asymptotic distribution.  All the previous \verb"Stata" commands provide functionality for unbalanced panels. 
\citet*{StammannHeissMacFadden2016} implemented the analytical correction in \verb"R" for estimators of parameters and APEs in logit and probit models with exogenous covariates and individual effects.

\section{EXTENSIONS}
\label{sec:extensions}

 \subsection{Unbalanced Panels}

In the previous section, we  assumed that the observations for all the combinations of the two indexes $i$ and $t$ were observed, i.e.  the panel was balanced. In empirical applications, however, it is common to have unbalanced panels where some of the observations are missing due to sample attrition.  This does not introduce special theoretical complications provided that the source of the missing observations is random. Perhaps due to this reason, we are not aware of any work in the panel literature that extends the bias corrections explicitly to unbalanced panels. To fill this void,  we provide without proof the asymptotic distribution of the FE for unbalanced panels.

Let ${\cal D}$ be the set of all observed pairs $(i,t)$, and $n = |{\cal D}|$ be the sample size. Define the operators $\mathbb{E}_n := n^{-1} \sum_{(i,t) \in {\cal D}}$,
$ \mathbb{E}_{T,i} = |{\cal D}_i|^{-1} \sum_{t \in {\cal D}_i}$,
and   $ \mathbb{E}_{N,t} = |{\cal D}_t|^{-1} \sum_{i \in {\cal D}_t}$, where  
 ${\cal D}_i = \{ t : (i,t) \in {\cal D}\}$ and ${\cal D}_t = \{ i : (i,t) \in {\cal D}\}$. Let $a_{it}$ be an attrition indicator for the observation $(i,t)$. The key regularity conditions that are needed to derive the bias expressions are (i) $|{\cal D}_i|/T \to c_i >0$ as $T \to \infty$ for all $i$; (ii) $|{\cal D}_t|/N \to c_t >0$ as $N \to \infty$ for all $t$; and (iii) $y_{it}$ is independent of $a_{it}$ conditional on $(x^t_i, \alpha, \gamma)$. The first two conditions guarantee that the number of observations for each unobserved effects increases with the sample size. The third condition imposes conditional missing at random in the attrition process.
The asymptotic distribution of $\widehat \beta$ then becomes
 \begin{equation*}
\widehat \beta - \beta_0 \overset{a}{\sim} {\cal N}\left( 
 \frac{B_u}{\overline T}     +  \frac{D_u}{\overline N}     , \; \; \frac{H_u^{-1}}{n}  \right), 
\end{equation*} 
where $H_u = - \plim_{n \rightarrow \infty} \mathbb{E}_n \overline \ell^{*\, \beta \beta}_{it}$, $\overline T = n / N$ is the average number of observations available for each cross-sectional unit, \begin{equation*}
B_u =   H_u^{-1}  \plim_{N,T \rightarrow \infty}
       \EN
        \left\{
         \frac{ -  \mathbb{E}_{T,i} \sum_{t \leq s \in {\cal D}_i}   \Ep \left( \ell^{\alpha_i}_{it}  
               \ell^{*\, \beta \alpha_i}_{is}  
                 \right)
                 -  \mathbb{E}_{T,i}   \overline  \ell^{*\, \beta \alpha_i \alpha_i}_{it} /2                
                 }
                              { \mathbb{E}_{T,i} \, \overline \ell^{\, \alpha_i \alpha_i}_{it} }   
          \right\}  ,         
\end{equation*}
$\overline N = n / T$ is the average number of observations available for each time period, 
 and 
\begin{equation*}
  D_u =    H_u^{-1}  \plim_{N,T \rightarrow \infty}
       \ET
        \left\{
         \frac{   -  \mathbb{E}_{N,t} \Ep \left( \ell^{\gamma_t}_{it}  
               \ell^{*\, \beta \gamma_t}_{it}  
                 \right)
                 -  \mathbb{E}_{N,t}  \overline  \ell^{*\, \beta \gamma_t \gamma_t}_{it} /2                
                 }
                              {   \mathbb{E}_{N,t} \, \overline \ell^{\, \gamma_t \gamma_t}_{it} }   
          \right\}     .
\end{equation*}
This result agrees with formula \eqref{BiasOrder}, because the order of the bias is $\overline T^{-1}   B  +  \overline N^{-1}  D   \sim (N+T)/n$. The magnitude of the bias depends on the averages $\overline T$ and $\overline N$, so having some units $i$ and $t$ with a very small number of observations is not necessarily a serious problem from
the perspective of incidental parameter bias. 

Regarding bias correction, the ABC for the case of unbalanced panels
 can be constructed using the empirical analog of the bias expressions evaluated at the FE estimator. Jackknife corrections can be formed by partitioning the panel in the same way as in the balanced case, i.e. without taking into account the attrition. We refer to \cite*{DhaeneJochmans2015}, \cite*{Stata2015},
\cite*{ChudikPesaranYang2016} and \cite*{Sun2016stata} for details.

\subsection{Multivariate Fixed Effects}
\label{sec:MultivariateFE}

We now briefly discuss the case where the unobserved effects are vector-valued, i.e. $\alpha_i \in \mathbb{R}^{d_{\alpha}}$ and $\gamma_t \in \mathbb{R}^{d_{\gamma}}$ with $d_{\alpha} \geq 1$ and $d_{\gamma} \geq 1$. We start by providing some motivating examples.

\paragraph*{Example (iv)}
An important case  is the normal  linear model with interactive effects or factor structure,
$y_{it} = x_{it}'\delta +  \alpha_{i}'\gamma_{t} + \varepsilon_{it},$ $\varepsilon_{it} \mid x^t_i, \alpha, \gamma, \beta   \sim \mathcal{N}(0,\sigma^2),$ where
\begin{equation*}\label{eq:linear2}
f(y \mid x_{it},\alpha_i, \gamma_t; \beta) = \frac{1}{\sqrt{2\pi \sigma^2}} \exp\left[-\frac{(y-x_{it}'\delta - \alpha_{i}'\gamma_{t} )^2}{2\sigma^2}\right], \quad \beta = (\delta,\sigma^2),   
\end{equation*}
Here, $d_{\alpha} = d_{\gamma}$ is the number of interactive effects.
\citet*{bai2009panel} showed that this model does not suffer from the incidental parameter problem when the covariates are strictly exogenous and the semiparametric model is correctly specified. However, he  found that  the FE least squares estimator  has an asymptotic bias structure of the form $B/T+D/N$ under conditional heteroskedasticity and derived the expressions for $B$ and $D$ that can be used for analytic bias correction.
 \citet*{MoonWeidner2017} showed that the FE least squares estimator of $\delta$ suffers from a \citet*{Nickell81} type incidental parameter bias when the covariates are predetermined, 
even when the model is correctly specified. We refer to \citet*{bai2016econometric} for a recent review,
 and to \citet*{Pesaran2006} and \citet*{ChudikPesaran2015} for alternative estimation methods
of factor models.

\paragraph*{Example (v)}
Another  example is a nonlinear single index factor model  where
\begin{equation*}
f(y \mid x_{it},\alpha_i, \gamma_t; \beta) = f(y,x_{it}'\beta +  \alpha_{i}'\gamma_{t}), \ \ d_{\alpha} = d_{\gamma}, 
\end{equation*} 
where $f$ is a known function such as $f(y,u) = F_{\varepsilon}(u)^y(1-F_{\varepsilon}(u))^{1-y}1(y \in \{0,1\})$ with a CDF $F_{\varepsilon}$, for a binary response model, or $f(y,u) = [\exp(u)^y \exp(-\exp u)/y!] 1(y \in \{0,1, ... \})$ for a count response model.  \citet*{CFW2014} characterized the bias of the FE estimator for models where the function $u \mapsto f(\cdot, u)$ is log-concave including the probit, logit, ordered probit and Poisson. See also  
\citet*{Chen2016estimation}, \citet*{BonevaLinton2017} and \citet*{Wang2017} for other articles discussing this model.

The analysis of Section~\ref{sec:SPM} carries over to the multivariate case with some minor adjustments. 
For the second-order
expansion of $\widehat \beta$ it is still convenient to define $\ell^*_{it}(\beta,\alpha_i,\gamma_t)$
as in \eqref{DefLstar}, but now $ \kappa_i$ is a $d_{\beta} \times d_{\alpha}$ matrix and
  $\rho_t$ is a $d_{\beta} \times d_{\gamma}$ matrix, which are solutions to
\begin{align*}
     \mathbb{E}_{T}  \left[ \overline \ell_{it}^{\, \beta \alpha_i'}  
                      +  \kappa_i  \,  \overline  \ell_{it}^{\, \alpha_i \alpha_i'}  
                      +  \rho_t \,  \overline  \ell_{it}^{\, \alpha_i \gamma_t'}  \right] &= 0 ,
          \qquad i=1,\ldots,N ,
     \\          
     \mathbb{E}_{N}  \left[ \overline \ell_{it}^{\, \beta \gamma_t'}  
                      +  \kappa_i  \,  \overline  \ell_{it}^{\, \alpha_i \gamma_t'}  
                      +  \rho_t \,  \overline  \ell_{it}^{\, \gamma_t \gamma_t'}  \right] &= 0 ,
          \qquad t=1,\ldots,T .
\end{align*}
Here we need to be more careful with the column and row dimensions of matrices in the notation. For example,
$\overline \ell_{it}^{\, \beta \alpha_i'}  := \partial(\partial \overline \ell_{it}/ \partial \beta)/ \partial \alpha_i'$ is the $d_\beta \times d_{\alpha}$ matrix
of second derivatives of $\overline \ell_{it}$.
A second-order asymptotic expansion similar to Section \ref{sec:AsBias} gives
\begin{equation*}\label{eq:adfe2}
\widehat \beta - \beta_0 \overset{a}{\sim} {\cal N}\left(   \frac B T + \frac D N, \frac{H^{-1} \Omega H^{-1}} {NT} \right),
\end{equation*} 
where $H = - \plim_{N,T \rightarrow \infty} \mathbb{E}_{NT} \overline \ell^{*\, \beta \beta'}$,
$\Omega =    \plim_{N,T \rightarrow \infty} \mathbb{E}_{NT} 
    \Ep\left( \overline \ell^{*\, \beta} \overline \ell^{*\, \beta'} \right)$,
 $B= H^{-1} b$ and $D = H^{-1} d$, and
$b$ and $d$ are $d_\beta$-vectors with elements
 \begin{align*}
    b_k  & := 
       \mathbb{E}_N
       \left\{     -    {\rm Tr}\left[
    \left(  \mathbb{E}_T \, \overline \ell^{\, \alpha_i \alpha_i'}_{it}  \right)^{-1}
        \mathbb{E}_T  \,
          \sum_{s=t}^T   \Ep \left( \ell^{\alpha_i}_{it}    \ell^{*\, \beta_k \alpha_i'}_{is} \right)  
           \right]
      \right.
      \\     
        & \qquad  \qquad
      \left.  
         + \frac 1 2    {\rm Tr}\left[
                                  \left( \mathbb{E}_T \, \overline \ell^{\, \alpha_i \alpha_i'}_{it}  \right)^{-1} 
                      \left( \mathbb{E}_T \,  \overline  \ell^{*\, \beta_k \alpha_i \alpha_i'}_{it}  \right)
                                  \left( \mathbb{E}_T \, \overline \ell^{\, \alpha_i \alpha_i'}_{it}  \right)^{-1} 
                         \mathbb{E}_T \,  \Ep \left( \ell^{\alpha_i}_{it}  \ell^{\alpha_i'}_{it}   \right)
                     \right]    
                      \right\}        ,  
\end{align*}
and
 \begin{align*}
    d_k  & := 
       \mathbb{E}_T
       \left\{     -    {\rm Tr}\left[
    \left(  \mathbb{E}_N \, \overline \ell^{\, \gamma_t \gamma_t'}_{it}  \right)^{-1}
        \mathbb{E}_N  \,
            \Ep \left( \ell^{\gamma_t}_{it}    \ell^{*\, \beta_k \gamma_t'}_{it} \right)  
           \right]
      \right.
      \\     
        & \qquad  \qquad
      \left.  
         + \frac 1 2    {\rm Tr}\left[
                                  \left( \mathbb{E}_N \, \overline \ell^{\, \gamma_t \gamma_t'}_{it}  \right)^{-1} 
                      \left( \mathbb{E}_N \,  \overline  \ell^{*\, \beta_k \gamma_t \gamma_t'}_{it}  \right)
                                  \left( \mathbb{E}_N \, \overline \ell^{\, \gamma_t \gamma_t'}_{it}  \right)^{-1} 
                         \mathbb{E}_N \,  \Ep \left( \ell^{\gamma_t}_{it}  \ell^{\gamma_t'}_{it}   \right)
                     \right]    
                      \right\}        .        
\end{align*}

In contrast to Section \ref{sec:AsBias}, the expressions of the asymptotic  bias and variance here  do not use the information equality to simplify terms, e.g.,  $H^{-1} \Omega H^{-1}= H^{-1}$. These formulas remain valid in conditional moment models where only a characteristic of the conditional distribution such as the expectation is correctly specified. This covers for example the linear model with interactive effects under heteroskedasticity, that is, 
a version of Example (iv) that only imposes
$$
\Ep[y_{it} \mid x_{it},\alpha_i, \gamma_t; \beta] = x_{it}'\beta +  \alpha_{i}'\gamma_{t}.
$$
The above formulas for asymptotic bias and variance thus contain those in 
\citet*{bai2009panel} and \citet*{MoonWeidner2017} as special cases.\footnote{See \citet*{Sun2016} for a related derivation.}  However, these results rely on an auxiliary consistency proof, which is usually model specific and therefore can be delicate. For example, it is  difficult to show consistency in models with interactive effects because the 
log-likelihood $(\beta,\alpha,\gamma) \mapsto \ENT \ell_{it}(\beta,\alpha_i,\gamma_t)$ is not concave.

Analytical and jackknife corrections can be formed analogously to Section \ref{subsec:bc}. 
For details we refer to \citet*{ArellanoHahn2016}   for models with individual effects,
\citet*{bai2009panel} and \citet*{MoonWeidner2017} for linear models with interactive fixed effects,
and to \citet*{CFW2014} for single-index factor models. \citet*{twostep11} derived bias corrections for two-step estimators of selection and other control variable models with unobserved individual effects in both steps. 
 \citet*{chudik2015common} and \citet*{Westerlund2018} developed bias corrections for the common correlated effects estimator of \citet*{Pesaran2006} for linear factor models. 

\subsection{Distributional and Quantile Effects}

 \citet*{CFW17} used nonlinear panel data methods to estimate distributional and quantile effects.  The idea is to model the distribution of the outcome conditional on the covariates and unobserved effects via distribution regression,  that is
$$
\Pr(y_{it} \leq y \mid x_{it}, \alpha_i, \gamma_t) = F_{y}(x_{it}'\delta(y) + \alpha_i'\pi(y) + \gamma_t' \xi(y) ), \ \ \beta(y) = (\delta(y),\pi(y),\xi(y)),
$$ 
where $F_{y}$ is a CDF such as the normal or logistic, $y \mapsto \beta(y)$ is a function-valued parameter, and the dimension of $\alpha_i$ and $\gamma_t$ is unrestricted. Both $F_y$ and $\beta(y)$ can vary with $y$ to accommodate heterogeneity accross the distribution.  At each $y$, the parameters are estimated by a binary response regression of the outcome $1(y_{it} \leq y)$ on the covariates and fixed effects with the parametrization $\alpha_i(y) = \alpha_i'\pi(y)$ and $\gamma_t(y) =  \gamma_t' \xi(y)$. The distributional and quantile effects are functionals of APEs. For example, the marginal distribution of the potential outcome corresponding to $x_{it}= x^0_{it}$ is the APE
$$
F^0(y) = (NT)^{-1} \sum_{i=1}^N\sum_{t=1}^T F_{y}(\delta(y)'x^0_{it} + \alpha_i(y) + \gamma_t(y) ),
$$ 
and the corresponding $\tau$-quantile is the functional
$$
Q^0(\tau) = \inf\{y \in \mathbb{R} : F^0(y) \geq \tau\}.
$$
The $\tau$-quantile effect of a change from $x_{it}= x^0_{it}$ to $x_{it}= x^1_{it}$ is
\begin{equation}\label{eq:qe}
Q^1(\tau) - Q^0(\tau),
\end{equation}
where $Q^1(\tau)$ is constructed using the same procedure as $Q^0(\tau)$ but with $x_{it}^1$ in place of $x_{it}^0$. 
\citet*{CFW17}  applied bias corrections to estimate and make inference on quantile effects uniformly over quantile indexes. These methods applied to continuous, discrete and mixed continuous-discrete outcomes.

The  effect in \eqref{eq:qe} is a marginal quantile effect that has causal interpretation under standard conditional ignorability conditions. Alternative conditional effects can be estimated by modelling the distribution of the outcome conditional on the covariates and unobserved effects via quantile regression. The FE quantile regression estimator,
introduced by  \citet*{Koenker2004},
 in general  suffers from the incidental parameter problem.
\citet*{KatoGalvaoMontes-Rojas2012} showed
asymptotic normality and unbiasedness of the FE quantile regression estimator in models with only individual effects under sequences where
 $N/T\rightarrow 0$ as $N,T \to \infty$.
\citet*{GalvaoKato2016smooth} derived the asymptotic distribution of the estimator in the same models, including the leading order
$1/T$ bias, for a smoothed version of the quantile regression objective function under sequences
for which $N$ and $T$ grow at the same rate.
\citet*{ArellanoWeidner2017} showed that the bias of the FE estimator without smoothing is the same as in \citet*{GalvaoKato2016smooth} in models with strictly exogenous regressors.
Thus, in terms of leading order bias and variance the FE quantile regression estimator behaves
analogous to any other non-linear panel data model. However,  the analysis in Section \ref{sec:SPM}  does not directly carry over to this case,
because the  quantile regression objective function is not sufficiently smooth.
We refer to  \citet*{GalvaoKato2018} for a recent review of quantile regression methods for panel data.

\subsection{Three Way Fixed Effects}

Another natural extension of the panel semiparametric model of Section~\ref{sec:SPM}
is a model with three-way effects for data with a multidimensional structure.   Let  
  $\ell_{ijt}( \beta, \alpha_i , \gamma_j, \delta_t) =   \log f(y_{ijt} \mid x_{ijt},  \alpha_i , \gamma_j, \delta_t;  \beta)$
be the log-likelihood of an outcome $y_{ijt}$ conditional on $x_{ijt}$, with common parameter $\beta$
and unobserved effects $\alpha_i$, $\gamma_j$ and $\delta_t$. Here, we have three panel indices
$i=1,\ldots,I$, $j=1,\ldots,J$ and $t=1,\ldots,T$. For example in an international trade application, \citet*{helpman2008estimating} are interested in analyzing the volume of trade from country $i$ to country $j$ at year $t$.
The corresponding fixed effect estimator is
\begin{align*}
  (\widehat \beta, \widehat \alpha,\widehat \gamma, \widehat \delta) & \in
  \argmax_{(\beta, \alpha, \gamma, \delta) \in \mathbb{R}^{d_{\beta}+I+J+T}} 
   \sum_{i=1}^I \sum_{j=1}^J \sum_{t=1}^T \ell_{ijt}( \beta, \alpha_i , \gamma_j, \delta_t) .
\end{align*}

\paragraph*{Example (vi)}
As an extension of Example (ii) above, consider a  binary response single index model 
$$y_{ijt} = 1(x_{ijt}'\beta + \alpha_i + \gamma_j + \delta_t \geq \varepsilon_{it}), \quad \varepsilon_{it} \mid x^t_i, \alpha, \gamma, \delta   \sim F_{\varepsilon},$$
where $F_{\varepsilon}$ is a known CDF.
For example, $y_{ijt}$ can be an indicator of trade from country $i$ to country $j$ at year $t$, $x_{ijt}$ the level of tariffs from country $i$ to country $j$ at year $t$, $\alpha_i$ is an exporter country effect, $\gamma_j$ is an importer country effect, and $\delta_t$ is a year effect.

Similar to the problem of unbalanced panels above, we are not aware of any formal discussion of three-way fixed effect models in the econometrics literature on large panels, but such estimation problems certainly occur in applications.
We provide a brief discussion here. In particular, we highlight the applicability
 of the formula \eqref{BiasOrder} to this case.
According to that formula, the expected order of the incidental parameter bias in $\widehat \beta$
 is $(I+J+T)/ (IJT)$, that is, we expect
a leading order bias structure $B_1/(IJ) + B_2/(IT) + B_3/(JT) $.
Compared to the order of the standard deviation of $\widehat \beta$, $1/\sqrt{IJT}$, the bias is small, as long
as all three panel dimensions are sufficiently large, in which case the incidental parameter problem can be ignored.
Thus, if $I,J,T \rightarrow \infty$ such that $I/JT \to 0$, $J/IT \to 0$ and $T/IJ \to 0$,
then $\widehat \beta$ is asymptotically normal and unbiased,
and standard MLE inference results are applicable to $\widehat \beta$,
without requiring any bias correction.

So far we have discussed a relatively benign three-way fixed effect model. 
A more difficult case is   $\ell_{ijt}( \beta,  \alpha_{ij} , \gamma_{it}, \delta_{jt}) =   \log f(y_{ijt} \mid x_{ijt},  \alpha_{ij} , \gamma_{it}, \delta_{jt};  \beta)$
and
\begin{align*}
  (\widehat \beta, \widehat \alpha,\widehat \gamma, \widehat \delta) & \in
  \argmax_{(\beta, \alpha, \gamma, \delta) \in \mathbb{R}^{d_{\beta}+IJ+IT+JT}} 
   \sum_{i=1}^I \sum_{j=1}^J \sum_{t=1}^T \ell_{ijt}( \beta,  \alpha_{ij} , \gamma_{it}, \delta_{jt}) .
\end{align*}
Here, the unobserved effects $\alpha_{ij}$,  $\gamma_{it}$ and $\delta_{jt}$ are pair-specific, only constant along
one of the panel dimensions. It is still possible to estimate them consistently if $I,J,T \rightarrow \infty$.
Applying the formula  \eqref{BiasOrder}, we expect the leading order incidental parameter bias to be of the form
$B_1/I + B_2/J + B_3/T$ since  $p=IJ+IT+JT$ and $n=IJT$. The order of this bias is never negligible compared to the order of the standard
deviation of $\widehat \beta$, $1/\sqrt{IJT}$. For example, if $I = J = T$, then the order of the  bias is $I^{-1}$, greater than the order of the standard deviation $I^{-3/2}$. It is therefore particularly useful to develop bias corrections for  models with fixed effects in multiple dimensions.

\section{CONCLUSION}
\label{sec:conclusions}
We have reviewed developments for large panel data models since \citet*{ArellanoHahn2007}.  A key insight is that the  order of the incidental parameter bias of the fixed effects estimators can be obtained from the simple formula \eqref{BiasOrder}. This formula  is useful to get an initial idea of the
relevance of the bias  in a given application, but the exact magnitude of the bias is still very much
model and data generating process dependent. In particular, it is impossible to give any general answer to the question of how large $T$ needs to be for the large-$T$ methods
to perform well, see for example \citet*{Greene2004} for simulation results on various non-linear panel data models.
We have discussed bias correction methods that estimate the actual size of the leading order bias. 
If this leading order bias is small (i.e. if the bias corrected estimator is close to the uncorrected estimator), then it is plausible to assume that the higher order bias terms 
are also small and can be neglected. 
If the leading order bias is  large (i.e. if the bias corrected estimator is very different from the uncorrected estimator),
then it is useful to conduct a Monte Carlo simulation  
calibrated to the application of interest, in order to verify that the remaining bias is small and that the inference
procedure works well.  

Fixed effect estimation methods for long panels
have recently been  applied to other data structures such as network and trade data. Examples include 
\citet*{Harrigan1996}, \citet*{AndersonWincoop2003}, 
\citet*{SantosSilvaTenreyro2006} and \citet*{helpman2008estimating}, which estimate gravity equations with unobserved importer and exporter country effects; and  \citet*{Graham2017},   \citet*{Dzemski2017}, 
 \citet*{Jochmans2017semiparametric}, \citet*{jochmans2017two},
\citet*{shi2016structural} and \citet*{Candelaria2016}, which apply large-$T$ panel methods to network data with unobserved sender and receiver effects. The large-$T$ panel methods are particularly well-suited for this type of data because typically $N=T$. Thus,
trade/network usually correspond to square panel data sets where the two dimensions index the same set of countries/individuals as senders and receivers. 

There are alternative semiparametric methods to FE that are not reviewed here. An interesting recent approach is the grouped fixed effects (GFE) of \citet*{hahn2010panel} and \citet*{bonhomme2015grouped} for models with individual effects. Compared to FE,  GFE is less affected by the  incidental parameter problem, because  it restricts the distribution of the individual effects to be discrete.
Large $T$ is still required to consistently estimate the group membership of individuals,
but if the true heterogeneity structure is indeed discrete, then 
no asymptotic bias correction is  required for the parameters of interest, even if $N$ is much larger than $T$.
The GFE approach imposes restrictions on the distribution of the individual effects, but  is more flexible in other dimensions than FE. For example, it permits to include individual effects that change over time. \citet*{bonhomme2017discretizing} characterized the approximation properties of GFE when the distribution of the individual effects is continuous. They showed that GFE is an effective dimension reduction device in this case, but suffers from the incidental parameter problem.

\section*{APPENDIX:  BIAS DERIVATION FOR PANEL MODELS}


We want to provide a heuristic derivation of the bias formulas~\eqref{eq:B} and \eqref{eq:D}.
Compared to section~\ref{sec:IIDexpansion},
one needs to define $\theta=(\beta, \alpha, \gamma)$, replace the observation index $j$ by the double index $it$,
and account for nonrandom sampling across $i$ and $t$. 
The appropriate generalization of the second-order expansion
in \eqref{SecondOrderExpansionIID} then reads
\begin{align}
  \label{SecondOrderExpansionPanel}
  \widehat \theta - \theta_0 &\approx  (NT)^{-1/2}  \psi_1 + (NT)^{-1}  \psi_2  ,
\end{align}
where
\begin{align*}
  \psi_1 & := - \left(  \mathbb{E}_{NT} \overline \ell^{\, \theta \theta}_{it}  \right)^{-1}  \GNT  \ell^\theta_{it}     ,
  \\ \nonumber
  \psi_2 &:= -   \left(  \mathbb{E}_{NT} \overline \ell^{\, \theta \theta}_{it}  \right)^{-1} \GNT  \ell^{\theta \theta}_{it}   \, \psi_1  
                     - \frac 1 2  \sum_{k=1}^p
                      \left(  \mathbb{E}_{NT} \overline \ell^{\, \theta \theta}_{it} \right)^{-1}  
                     \left( \mathbb{E}_{NT}  \overline \ell^{\theta \theta \theta_k}_{it} \right)  \, \psi_1 \psi_{1,k} .
\end{align*}  
Here, the expected Hessian $\Ep  \ell^{\theta \theta}_{j}$ in \eqref{SecondOrderExpansionIID}
was replaced with the sample average of the expected Hessian 
$ \mathbb{E}_{NT} \overline \ell^{\, \theta \theta}_{it} = (NT)^{-1} \sum_{i,t} \Ep \ell^{\, \theta \theta}_{it} $,
and analogously for the third derivative term $ \Ep  \ell^{\theta \theta \theta_k}_{j} $.
Similarly, we need to define 
$\GNT  \ell^{\theta \theta}_{it} $ here
as $\GNT  \ell^{\theta \theta}_{it} 
   = (NT)^{-1/2} \sum_{i,t} \left(   \ell^{\theta \theta}_{it} -  \overline \ell^{\theta \theta}_{it} \right) $.
Apart from those changes  the derivation in section~\ref{sec:IIDexpansion} still applies.
One always first needs an additional consistency argument for $  \widehat \theta$
before the expansion~\eqref{SecondOrderExpansionPanel} becomes applicable.

\subsection*{One-way Fixed Effects}

We start by considering
the estimation problem \eqref{eq:fe} with only individual specific effects,
that is, $(\widehat \beta,\widehat \alpha) = \argmax_{(\beta,\alpha) \in \mathbb{R}^{d_{\beta} + N}}  \,  \mathbb{E}_{NT}  \ell_{it}(\beta, \alpha_i)$.
It is convenient to define 
\begin{align}
    \ell^*_{it}(\beta,\alpha_i)  &:=  \ell_{it}(\beta, \alpha^*_i(\beta,\alpha_i)) ,
    &
     \alpha^*_i(\beta,\alpha_i) &= \alpha_i 
        - \left( \mathbb{E}_T \overline \ell^{\, \alpha_i \alpha_i}_{it} \right)^{-1}
     \left(  \mathbb{E}_T \overline \ell^{\, \beta \alpha_i }_{it} \right)'  \beta .
   \label{Reparameterize1}  
\end{align} 
The profile objective functions $ \max_{\alpha}  \,  \mathbb{E}_{NT} \ell_{it}(\beta, \alpha_i) $
and   $ \max_{\alpha}  \,  \mathbb{E}_{NT}  \ell^*_{it}(\beta,\alpha_i) $ have the same maximizer
$\widehat \beta$,
because we have simply re-parameterized $\alpha_i$.
 $\mathbb{E}_{NT} \ell^{*}_{it}(\beta,\alpha_i)$ is information-orthogonal between $\beta$ and $\alpha_i$, i.e.  $\mathbb{E}_{NT} \overline \ell^{* \, \beta \alpha_i}_{it} = 0$ for all $i = 1,\ldots,N$,
where the omitted parameter means evaluation at the true parameters after the parameter transformation.\footnote{
We can express partial derivatives of $ \ell^*_{it}(\beta,\alpha_i)$ in terms of partial derivatives of 
$ \ell_{it}(\beta,\alpha_i)$, for example, 
$\ell^{* \,  \beta  \alpha_i}_{it}(\beta,\alpha_i)
 = \ell^{\beta  \alpha_i}_{it}(\beta,\alpha^*_i(\beta,\alpha_i))
    -  \ell^{\alpha_i  \alpha_i}_{it}(\beta,\alpha^*_i(\beta,\alpha_i))
      \left( \mathbb{E}_T \overline \ell^{\, \alpha_i \alpha_i}_{it} \right)^{-1}
     \left(  \mathbb{E}_T \overline \ell^{\, \beta \alpha_i}_{it} \right)$.
From this we also immediately find that $\mathbb{E}_T \overline \ell^{* \,  \beta  \alpha_i}_{it} = 0$,
which also implies $\mathbb{E}_{NT} \overline \ell^{* \, \beta \alpha_i}_{it} = 0$.
}  
Thus, the expected Hessian
of the objective function $\mathbb{E}_{NT}  \ell^*_{it}(\beta,\alpha_i) $ is a block-diagonal matrix,
with one $d_{\beta} \times d_{\beta}$ block $\mathbb{E}_{NT} \overline \ell^{* \, \beta \beta}_{it}$,
and the rest of the matrix diagonal.
By applying the expansion \eqref{SecondOrderExpansionPanel} to $\mathbb{E}_{NT}  \ell^*_{it}(\beta,\alpha_i) $
with $\theta=(\beta, \alpha)$ we find
\begin{align}
    \label{SecondOrderExpansionPanelBeta}
  \widehat \beta - \beta_0 &\approx  (NT)^{-1/2}  \psi_{1,\beta} + (NT)^{-1}  \psi_{2,\beta}  ,
\end{align}
with
\begin{align*}
  \psi_{1,\beta} & =  H_{NT}^{-1}  \GNT  \ell^{* \, \beta}_{it}     ,
&
  \psi_{2,\beta} &\approx   H_{NT}^{-1}  
       \sum_{i=1}^N   
         \left[  
         -  \frac{  \left(  \mathbb{G}_{T}  \ell^{* \, \beta \alpha_i}_{it}  \right)
                      \mathbb{G}_{T}   \ell^{\alpha_i}_{it}    }
                   {    \mathbb{E}_{T}  \overline \ell^{\, \alpha_i  \alpha_i}_{it}     }        
                      + 
                      \frac{   \left(   \mathbb{E}_{T}  \overline \ell^{* \, \beta \alpha_i \alpha_i}_{it} \right)
                      \left(   \mathbb{G}_{T}   \ell^{\alpha_i}_{it} \right)^2 }
                      {2 \left(  \mathbb{E}_{T} \overline \ell^{\, \alpha_i  \alpha_i}_{it} \right)^2 } \right] ,
\end{align*}  
where  $H_{NT} =  - \mathbb{E}_{NT} \overline \ell^{*\, \beta \beta}_{it}$. In $\psi_{2,\beta}$ we have dropped the terms that would originate from $\GNT  \ell^{\beta \beta}_{it} $,
 $ \overline \ell^{\beta \beta \beta}_{it}  $ and $ \overline \ell^{\beta \beta \alpha_i}_{it}  $, because they only give smaller order terms. We also used that partial derivatives with respect to only $\alpha_i$ are equal for $\ell_{it}$ and $\ell^*_{it}$.
The leading bias of order $T^{-1}$ in $\widehat \beta$ is thus given by
$N^{-1} \Ep \psi_{2,\beta} \approx B_{NT}$, where
 \begin{align}
    B_{NT}  := H_{NT}^{-1}
       \mathbb{E}_N
        \left[
       -  \frac{  \mathbb{E}_T  \,
          \sum_{s=t}^T   \Ep \left( \ell^{\alpha_i}_{it}  
               \ell^{*\, \beta \alpha_i}_{is}  
                 \right)
       }
                              {  \mathbb{E}_T \, \overline \ell^{\, \alpha_i \alpha_i}_{it} }
                     + 
                      \frac{ \left( \mathbb{E}_T \,  \overline  \ell^{*\, \beta \alpha_i \alpha_i}_{it}  \right)
                         \mathbb{E}_T \,  \Ep \left( \ell^{\alpha_i}_{it}  \right)^2   }  
                         {  2 \, \left( \mathbb{E}_T \, \overline \ell^{\, \alpha_i \alpha_i}_{it}  \right)^2 }
          \right]        .   
    \label{DefB0}         
\end{align}
 The last formula is also valid for conditional moment models, where the 
 information equality $ \Ep \left( \ell^{\alpha_i}_{it}  \right)^2  = -   \overline \ell^{\, \alpha_i \alpha_i}_{it}$
 may not hold. Assuming a correctly specified likelihood and applying the  information equality 
 gives the result for $B = \plim_{N,T \rightarrow \infty}   B_{NT}$ in equation \eqref{eq:B}.

\subsection*{Two-way Fixed Effects}

We now include the time effects as well, as in \eqref{eq:fe},
that is, we use the expansion \eqref{SecondOrderExpansionPanel}  with $\theta=(\beta, \alpha, \gamma)$.
Some normalization of $\alpha$ and $\gamma$
may be required,
and this can also result in the
Hessian matrix $\mathbb{E}_{NT} \overline \ell^{\, \theta \theta}_{it}$ to be singular, which needs
to be accounted for in   \eqref{SecondOrderExpansionPanel} by, for example, 
applying a pseudo-inverse instead of the regular inverse,
but for our heuristic discussion here this is not important.

Consider  $  \ell^*_{it}(\beta,\alpha_i,\gamma_t)$
as defined in \eqref{DefLstar}.
This definition of $\ell^*_{it}$
 guarantees that the expected Hessian matrix of $\mathbb{E}_{NT}   \ell^*_{it}(\beta,\alpha_i,\gamma_t) $
is a block-diagonal matrix when evaluated at the true parameters,
with  two non-zero blocks   $\mathbb{E}_{NT}   \overline \ell^{*  \, \beta \beta}_{it} $ 
and $\mathbb{E}_{NT}   \overline \ell^{*  \, \phi \phi}_{it} $, where $\phi=(\alpha,\gamma)$.
Here, $\mathbb{E}_{NT}   \overline \ell^{*  \, \phi \phi}_{it} $
is not a diagonal matrix, but its diagonal elements dominate, 
and in \cite*{FernandezValWeidner2016} it is shown that its (pseudo-) inverse 
 can be approximated by the inverse of its 
diagonal part.
The expansion in \eqref{SecondOrderExpansionPanelBeta}
is thus almost unchanged, we just need to add the terms in $\psi_{2,\beta}$ that stem from summing over
the parameters $\gamma_t$ as well,
\begin{align*}
  \widehat \beta - \beta_0 &\approx  (NT)^{-1/2}  \psi_{1,\beta} + (NT)^{-1}  \psi_{2,\beta}  ,
\end{align*}
where
\begin{align*}  
  \psi_{1,\beta}  =    H_{NT}^{-1}  \GNT  \ell^{* \, \beta}_{it}     ,
  \qquad
  \psi_{2,\beta} &\approx H_{NT}^{-1} \sum_{i=1}^N   
         \left[ - 
          \frac{    \GT   \ell^{* \, \beta \alpha_i}_{it} \GT  \ell^{\alpha_i}_{it}    }
                   {   \ET \overline \ell^{\, \alpha_i  \alpha_i}_{it}     }        
                      + 
                      \frac{   \ET \overline \ell^{* \, \beta \alpha_i \alpha_i}_{it} 
                      \left( \GT  \ell^{\alpha_i}_{it} \right)^2 }
                      {2 \left( \ET \overline \ell^{\, \alpha_i  \alpha_i}_{it} \right)^2 } \right]
           \\ \nonumber & \quad
            + H_{NT}^{-1}
              \sum_{t=1}^T
         \left[
         -  \frac{  \GN \ell^{* \, \beta \gamma_t}_{it} 
                     \GN  \ell^{\gamma_t}_{it}     }
                   {   \EN \overline \ell^{\, \gamma_t  \gamma_t}_{it}     }      
                      + 
                      \frac{   \EN  \overline \ell^{* \, \beta \gamma_t \gamma_t}_{it} 
                      \left(  \GN  \ell^{\gamma_t}_{it} \right)^2 }
                      {2 \left( \EN \overline \ell^{\, \gamma_t  \gamma_t}_{it} \right)^2 }     
                      \right]        ,
\end{align*}  
with $H_{NT} =  - \mathbb{E}_{NT} \overline \ell^{*\, \beta \beta}_{it}  $.
We   again dropped  terms from $\psi_{2,\beta}$ that are asymptotically irrelevant.
We now find
$(NT)^{-1}  \Ep \psi_{2,\beta} \approx T^{-1} B_{NT} + N^{-1} D_{NT}$,
where the formula for $B_{NT}$ in \eqref{DefB0} is unchanged,
and the time effects incidental parameter bias reads
\begin{align*}
     D_{NT}  &=  H_{NT}^{-1}
       \mathbb{E}_T
        \left[
       -  \frac{  \mathbb{E}_N  \,  \Ep \left( \ell^{\gamma_t}_{it}  
               \ell^{*\, \beta \gamma_t}_{it}  
                 \right)}
                              {  \mathbb{E}_N \, \overline \ell^{\, \gamma_t \gamma_t}_{it} }
                     + 
                      \frac{ \left( \mathbb{E}_N \,  \overline  \ell^{*\, \beta \gamma_t \gamma_t}_{it}  \right)
                         \mathbb{E}_N \,   \Ep\left( \ell^{\gamma_t}_{it}  \right)^2   }  
                         {  2 \, \left( \mathbb{E}_N \, \overline \ell^{\, \gamma_t \gamma_t}_{it}  \right)^2 }
          \right]        ,
\end{align*}
where we have not used the information equality, yet, so that the formula  holds
for conditional moment models as well.
Using $ \Ep\left( \ell^{\gamma_t}_{it}  \right)^2   = -  \overline \ell^{\, \gamma_t \gamma_t}_{it} $
then gives the result for  $D = \plim_{N,T \rightarrow \infty}   D_{NT}$ in equation \eqref{eq:D} of the main text.


\end{document}